\documentclass[10pt,aps,prl,twocolumn,superscriptaddress,reprint,balance]{revtex4-1} 

\usepackage{amssymb,amsmath}
\usepackage{graphicx}
\usepackage{xcolor}
\usepackage[normalem]{ulem} 
\usepackage{hyperref,tikz,pgfplots}
\usepackage{etoolbox}
\usepackage{orcidlink}
\graphicspath{{figures/}}
\usepackage[normalem]{ulem}
\usepackage{soul}

\sethlcolor{yellow}

\begin{document}


\title{Thin-film drainage becomes singular at saddles }

\author{Simeon Djambov\,\orcidlink{0000-0003-0979-1542}}
\affiliation{Laboratory of Fluid Mechanics and Instabilities, EPFL, Lausanne CH-1015, Switzerland}
\affiliation{LadHyX, CNRS, Ecole Polytechnique, Institut Polytechnique de Paris, Palaiseau, France}

\author{Alice Marcotte\,\orcidlink{0009-0001-1769-0621}}
\affiliation{Institut Jean le Rond d'Alembert, CNRS, Sorbonne Université, Paris, France}

\author{Fran\c{c}ois Gallaire\,\orcidlink{0000-0002-3029-1457}}
\affiliation{Laboratory of Fluid Mechanics and Instabilities, EPFL, Lausanne CH-1015, Switzerland}

\author{Pier Giuseppe Ledda\,\orcidlink{0000-0003-4435-8613}}
\email{piergiuseppe.ledda@unica.it}
\affiliation{Department of Civil, Environmental Engineering and Architecture, University of Cagliari, 09123 Cagliari, Italy}

\date{\today}

\begin{abstract}

{
From coating curved parts to creeping lava, thin-film gravitational drainage over complex surfaces can produce abrupt accumulation, usually associated with sharp edges or defects. Here, experiments, simulations and asymptotics show that a smooth saddle alone generates a logarithmically singular ridge, regularized within a dynamic region where drainage balances hydrostatic pressure and capillarity. In a Newtonian lubrication picture, this mechanism can inform redistribution in manufacturing and geophysics, motivating extensions to complex rheology and interacting geometric features.
}

\end{abstract}

\maketitle
Viscous drainage of thin liquid films over complex topographies arises in settings {spanning coating and manufacturing processes \cite{scriven1988physics,weinstein2004coating}, geophysical processes} \cite{neufeld2011leakage} {and} deposition and erosion in geological systems \cite{huppert1986intrusion,meakin2010geological,bertagni2021hydrodynamic}{. In these systems, substrate geometry can generate strong local variations in film thickness. Such variations affect coating in glove manufacturing and shell uniformity in investment casting \cite{dong2024shell} and control lava propagation and subsequent deposition at geophysical scales \cite{saville2022safe,younger2019marcath}. Identifying the geometric features that preferentially accumulate liquid is therefore relevant across a broad range of scales and applications.}

\begin{figure}[t]
    \centering
\includegraphics[width=\linewidth]{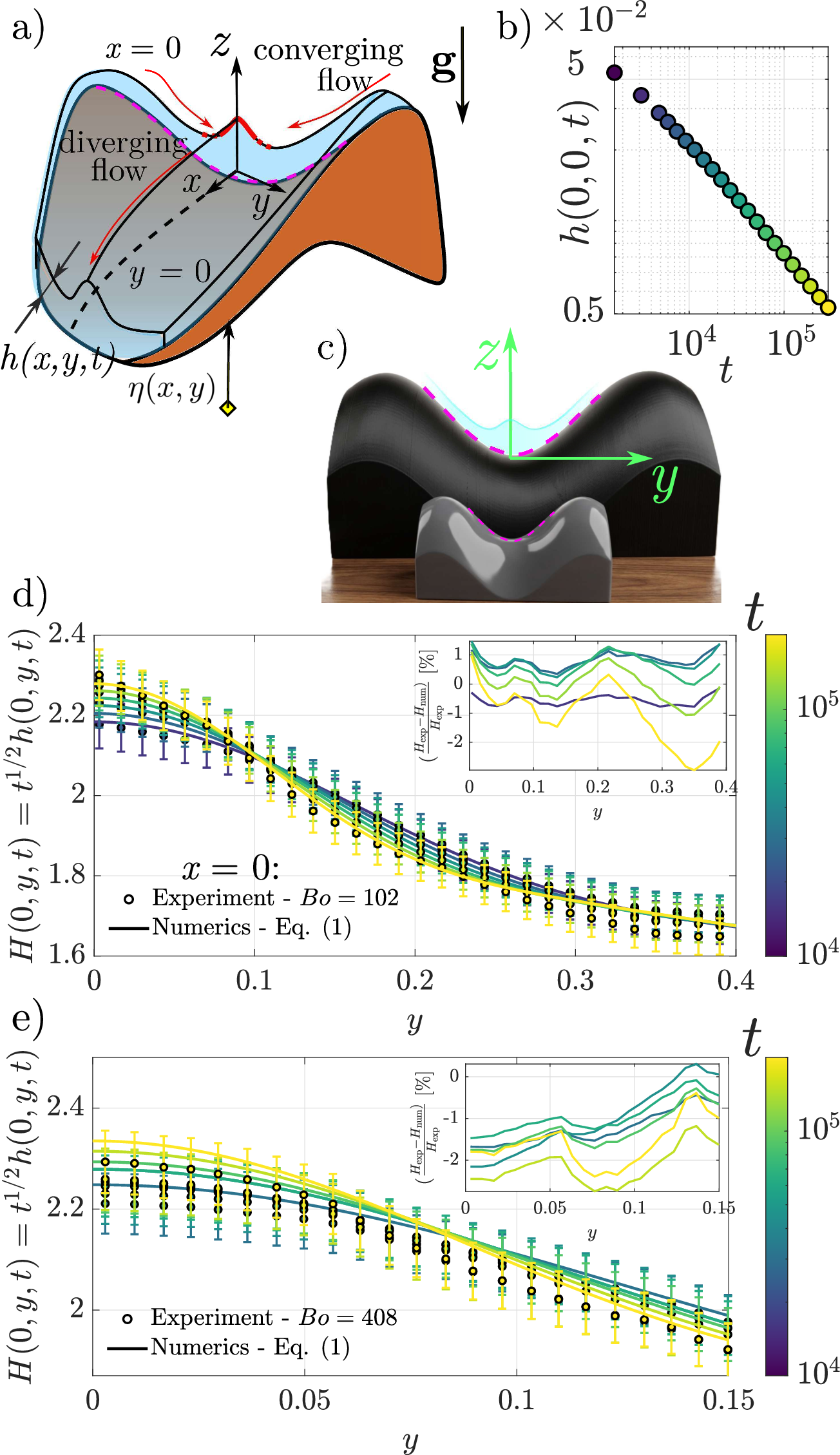}
    \caption{(a) Sketch of the drainage of a thin film over a saddle-shaped substrate. Liquid drains from elevated regions toward lower regions, producing converging inflow toward the saddle and diverging outflow away from it. (b) Experimental saddle thickness as a function of time, for the $\nu=10^{-3}\,$m$^2$s$^{-1}$ silicone oil, and $Bo=10{2}$. 
    (c) Experimental photo of {two of the three} employed substrates with sketch of the imaged thickness along $x=0$. (d,e) {Rescaled thickness profiles $H=t^{1/2}h$ along $x=0$ for
    (d) $Bo=102$ and (e) $Bo=408$.
    Circles denote experiments and solid lines numerical solutions of
    Eq.~\eqref{eq:fluxform} for {initial nondimensional thickness} $h_0=0.1$; colors indicate increasing
    time. Insets show the relative experiment--simulation difference
    $(H_{\mathrm{num}}-H_{\mathrm{exp}})/H_{\mathrm{exp}}$.}
    }
    \label{fig:1}
\end{figure}

In the viscous thin-film limit, the dynamics is governed by gravity, viscosity, substrate curvature, and capillarity \cite{oron1997long,roy2002lubrication,lee2016fabrication,balestra2019fingering,ledda2022gravity}.
Classical drainage is relatively well understood on flat substrates \cite{jeffreys1930draining,huppert1982flow,lister1992viscous,xue2020self} and on simple curved geometries such as cylinders, cones, spheres, and funnels. In these cases, drainage is predominantly converging or diverging {under the action of gravity along the surface, whose variation, set by the substrate curvature, determines the film thickness distribution} \cite{takagi2010flow,xue2021draining}. Much less is known when these two tendencies coexist locally. {Specifically, the signs of the principal curvatures determine the local topology of gravity-driven drainage. Opposite-sign curvatures, occurring in a saddle, produce a hyperbolic flow} with drainage locally converging along one principal direction and diverging along the other [Fig.~\ref{fig:1}]. {Sharp features are known to organize thin-film and interfacial
dynamics, including defects
\cite{perrin2016defects}, contact lines
\cite{he2019characteristic}, and edges
\cite{xue2020self}. This raises the question of whether smooth interior features such as saddles, which occur naturally on surfaces with nonuniform curvature, can similarly organize thin-film drainage.}

In this Letter, we combine experiments, numerical simulations, and theory to show that {the hyperbolic drainage topology near a smooth saddle focuses the film into a localized ridge whose leading-order profile is logarithmically singular,} without requiring edges, corners, contact lines, or imposed defects. The apparent singularity {induced by the hyperbolic flow} is regularized within a region where drainage, hydrostatic pressure, and capillarity balance. Its width is not set by a pre-existing geometric length, but is selected by the flow and shrinks as drainage proceeds.

We test  this mechanism in a controlled setting using {a Newtonian fluid on} a smooth substrate containing a single nondegenerate saddle. Lengths are scaled by a characteristic length \(L\), and the substrate is prescribed as
$
\hat{\eta}(\hat{x},\hat{y})=L-L\cos(\hat{y}/L)-\frac{\hat{x}^2}{2L}.
$
Near the origin, this surface reduces to
$
\hat{\eta}\simeq \frac{\hat{y}^2}{2L}-\frac{\hat{x}^2}{2L},
$
so that the saddle has opposite curvatures in the two, orthogonal, principal directions $\hat{x}$ and $\hat{y}${, and gravity acts downward along the $\hat{z}$ direction} [Fig.~\ref{fig:1}]. {The periodic dependence in $\hat{y}$ introduces neighboring maxima that delimit the drainage basin feeding the central saddle, thereby uniquely fixing its liquid supply and ensuring reproducible late-time drainage, independent of the initial conditions \cite{lee2016fabrication}. The experiment focuses on a minimal feeding unit, a central saddle surrounded by two maxima. Near the saddle, the local quadratic form controls the singular behavior derived below.}
{A silicone oil (Carl Roth) of density $\rho=970 \pm 5 \,$kg\,m$^{-3}$, kinematic viscosity $1000 \pm 50$  mm$^2\,$s$^{-1}$ (from manufacturer's datasheet), and surface tension $\gamma = 21 \pm 0.2\,$mN$\,$m$^{-1}$ \cite{wang2026} was used in the experiments. 
Three geometrically similar substrates reproducing the prescribed saddle 
geometry were fabricated by 3D printing, as detailed in the Supplemental Material \footnote{See Supplemental Material for further detail, which includes Refs. \cite{ledda2022gravity,bers1964partial,roy2002lubrication,howell2003surface,roberts2006accurate,thiffeault2006transport,wray2017reduced,shepherd2024general,irgens2019tensor}}. Their length scales were 
$L=15 \pm 0.025$, $30 \pm 0.1$, and $40 \pm 0.1\,\mathrm{mm}$. The corresponding Bond numbers, measuring the effect of gravity against surface tension, are
$Bo=\rho gL^2/\gamma=102\pm1$, $408\pm5$, and $725\pm 9$, respectively. 
}

The draining film was imaged with a uniform LED panel placed behind the substrate, while a camera equipped with a macro lens 
recorded the transmitted light from
the opposite side, along the plane $x=0$ [Fig.~\ref{fig:1}]. The thickness profile $\hat{h}(\hat{x}=0,\hat{y},\hat{t})$ was extracted by subtracting the dry-substrate reference image before coating {\cite{Note1}}. 
{The uncertainty on the measured film thickness $\hat h$ is primarily associated 
with the calibration factor converting pixel distances into physical lengths. 
This calibration was obtained by fitting the dry-substrate profile with the 
prescribed parabolic geometry. The resulting uncertainty on 
the calibration factor is below $\pm 2.5\%$\cite{Note1}.} 

Lengths are nondimensionalized by $L$, so that the nondimensional film thickness is $h=\hat h/L$, and time by the drainage scale $\tau_d=\nu/(gL)$ such that $t=\hat{t}/\tau_d$, where $g$ is the gravitational acceleration. {The experimental observations span the nondimensional time range 
$10^{4}\lesssim t \lesssim 5\times10^{5}$.} The measured saddle thickness decays with time [Fig.~\ref{fig:1}(b)]. Motivated by the standard late-time drainage law, we rescale the thickness as $H:=t^{1/2}h=(g\hat t/\nu L)^{1/2} \hat h $, independent of the thickness initial condition \cite{lee2016fabrication}. {The uncertainty on the rescaled thickness $H$ thus includes also the contribution from viscosity. 
}
{Although the thickness appears to follow  the $t^{-1/2}$ scaling, as shown in Fig.~\ref{fig:1}(b) for $Bo=102$, the rescaled profiles retain a time dependence in the immediate vicinity of the saddle, exhibiting a maximum that slightly grows over time [Fig.~\ref{fig:1}(d,e)].}

We model {viscous} drainage in the lubrication limit by a conservation law written in the natural coordinate system of the curved substrate \cite{ledda2022gravity},
\begin{equation}
\mathcal{V}\partial_t h+\nabla\cdot\left(\mathbf q_d+\mathbf q_r+\frac{1}{Bo}\mathbf q_\gamma\right)=0.
\label{eq:fluxform}
\end{equation}
Here $\mathbf q_d=\frac{1}{3}h^3 \mathbf{g}_t$ is the leading gravity-driven drainage flux, where $\mathbf{g}_t$ is tangential gravity; $\mathbf q_r$ collects the next-order hydrostatic-pressure flux corrections, and $\mathbf q_\gamma$ is the capillary flux. The factor $\mathcal{V}$ is a thickness-dependent geometric correction and reduces to unity as the film drains, $\mathcal{V}=1+\mathcal{O}(t^{-1/2})$. Explicit expressions for \(\mathcal{V}\), \(\mathbf q_r\), and \(\mathbf q_\gamma\) are reported in the Supplemental Material \cite{Note1}.
Eq.~\eqref{eq:fluxform} is solved via a custom formulation in COMSOL, previously validated against experiments \cite{ledda2022gravity}, with further detail reported in the Supplemental Material \cite{Note1}. 
{The numerical profiles reproduce experimental measurements, as quantified also by the relative difference in the insets {[Fig.~\ref{fig:1}d,e]}. The deviations are a few percent, in line with the experimental uncertainties.} 

\begin{figure}[!t]
    \centering
    \includegraphics[width=0.9\linewidth]{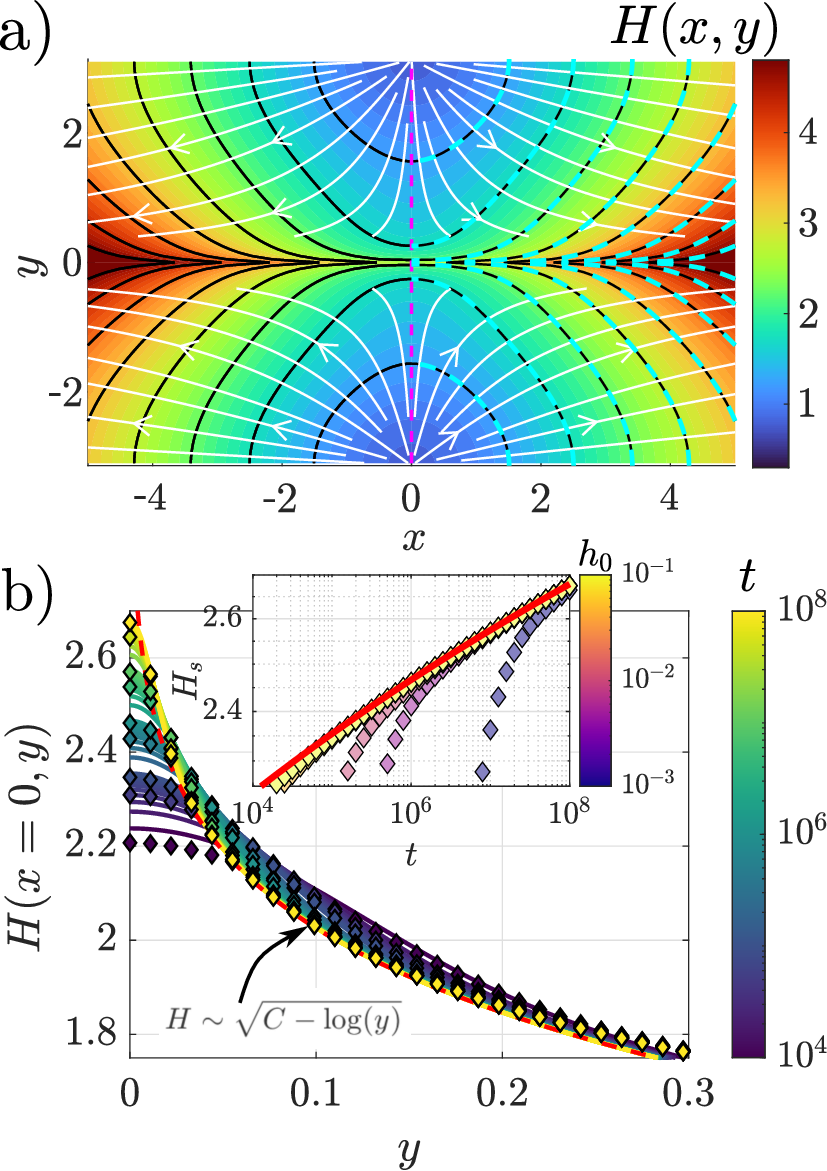}
    \caption{{Capillary-free problem. (a)} Late-time drainage field for $h(x,y,0)=0.1$ at $t=10^8$. Colormap and isocontours (black) of the numerical rescaled thickness $H(x,y)$. Cyan dashed lines show the drainage solution {$H_d$} [Eq.~\eqref{eq:outerchar}], and white curves denote the numerical drainage flux $\frac13 h^3\mathbf{g}_t$ streamlines. 
    (b) {Rescaled thickness profiles for increasing times. Diamonds are numerical solutions $H(x=0,y)$, colored solid lines are {theoretical} solutions ${\mathcal{H}}(x=0,y)$ {obtained by solving the saddle layer problem Eq.~\eqref{eq:innerhydro}}, and the red dashed line is the near-saddle drainage solution {$H_d^*$} with $C\approx 1.8$. Inset: numerical rescaled saddle thickness $H_s:=H(0,0)$ versus time, for different initial uniform thicknesses $h_0$. The theory (red solid line) captures both the ridge profile and the slow growth of $H_s$, independently of the initial condition.}}
    \label{fig:2}
\end{figure}

We first consider the late-time drainage problem for $Bo\to\infty$. Writing 
$
h(x,y,t)=t^{-1/2} H_d(x,y)+o(t^{-1/2}),\, t^{-1/2} \ll 1,
$
the $O(1)$ balance in Eq.~\eqref{eq:fluxform} retains only the leading drainage flux and yields \cite{lee2016fabrication,ledda2022gravity}
\begin{equation}
-\frac{H_d}{2}+\frac{1}{3}\nabla\cdot\left(H_d^3\mathbf g_t\right)=0.
\label{eq:outerflux}
\end{equation}
{where
$
\mathbf g_t=-\frac{\nabla_{xy}\eta}{w^2},\,
w=\left(1+\eta_x^2+\eta_y^2\right)^{1/2},
\label{eq:gtmonge}
$
}
{with} $\nabla_{xy}=[\partial_x,\partial_y]$, from which it follows $\nabla\cdot\mathbf g_t=-\mathcal{K}/w$, where $\mathcal{K}$ is twice the classical mean substrate curvature \cite{ledda2022gravity}. Eq.~\eqref{eq:outerflux} becomes
\begin{equation}
\nabla_{xy}\eta\cdot\nabla_{xy}H_d^2+w\left(\frac{2}{3}\mathcal{K}H_d^2+w\right)=0,
\label{eq:outerchar}
\end{equation}
solved by the method of characteristics \cite{Note1} with initial conditions imposed at the substrate maxima \cite{ledda2022gravity}. 
{Fig.~\ref{fig:2} compares the drainage solution $H_d$ with numerical
solutions of Eq.~\eqref{eq:fluxform} without capillarity. The geometry produces inflow along $y$ and outflow along $x$. Away from the saddle, $H=t^{1/2}h$ approaches $H_d$, whereas near $y=0$ a time-dependent ridge forms.} {Its} height and lateral extent remain time-dependent [{diamonds in} Fig.~\ref{fig:2}{b}], even though the rescaled thickness has already collapsed onto the drainage profile [Eq.~\eqref{eq:outerchar}], away from $y=0$. Conversely, the drainage solution $H_d$ grows without bound as one approaches $y=0$.

This behavior follows from the local saddle geometry.
Near $(x,y)=(0,0)$, the substrate is approximated by
$
\eta(x,y)\approx \frac{y^2}{2}-\frac{x^2}{2},
$
so that
$
\eta_x  \approx -x,\,
\eta_y  \approx y,\,
w  \approx 1,
$
while $\mathcal{K} \approx 0$. Eq.~\eqref{eq:outerchar} thus reduces to
$
-x\partial_x (H_d^2)+y\partial_y (H_d^2)+1=0.
$
Along the symmetry line $x=0$, where $\partial_x H_d=0$, this gives
\begin{equation}
\frac{dH_d^2}{dy}=-\frac{1}{y},
\implies
H_d(0,y)=\left(C-\log |y|\right)^{1/2}:=H_d^*(y){.}
\label{eq:outerlog}
\end{equation}
{The constant $C$} is set by {Eq.~\eqref{eq:outerchar}} by solving the drainage problem along the line $x=0$, as detailed in the Supplemental Material \cite{Note1}. {For the global (far from the saddle) substrate topology under consideration, $C \approx 1.8$.}
Thus, unlike previously studied geometries, the saddle forces a logarithmic divergence of the leading-order late-time profile, which follows from the local hyperbolic geometry $\eta \sim y^2-x^2$. 
The singularity is not caused by a sharp edge, contact line, or imposed defect, but by the structure of the gravitational drainage field near a smooth saddle. 

The singularity can be regularized without reference to the global shape of the substrate, by introducing a minimal local model that includes hydrostatic effects and, for completeness, capillary terms on flat surfaces \cite{oron1997long,xue2020self}. The metric is thus taken as Cartesian, and the leading-order effect of the saddle geometry is retained through the tangential gravity, 
$
\mathbf g_t\approx x\mathbf e_x-y\mathbf e_y.
$
Introducing the local rescaled thickness $\mathcal{H}$, $
h(x,y,t)=\delta \mathcal{H}(x,y)+o(\delta),\, \delta:=t^{-1/2} \ll1
$.
The local thin-film equation reads as \cite{oron1997long}
\begin{multline}
-\frac{\mathcal{H}}{2}
+\nabla_{xy}\cdot\left[
\frac{\mathcal{H}^3}{3}{\bigg(}
x\mathbf e_x-y\mathbf e_y  \right. \\ \left. \left.
-\delta\nabla_{xy}\mathcal{H}
+\frac{\delta}{Bo}\nabla_{xy}\nabla_{xy}^2 \mathcal{H}
\right)
\right]=0.
\label{eq:localmodeldimless}
\end{multline}
We begin by considering the case without capillarity ($Bo\to\infty$). Compared with the drainage problem, this local balance retains the same saddle-induced drainage flux but also includes the hydrostatic-pressure correction, \(-\delta \mathcal{H}^3\nabla_{xy}\mathcal{H}\), which becomes important as the drainage solution develops large gradients. We thus seek a narrow region localized in the transverse direction by introducing the stretched coordinate
$
y=\delta^\alpha\zeta.
$
In Eq.~\eqref{eq:localmodeldimless}, the drainage contribution along $y$ scales as $\delta^\alpha$, because it is proportional to $y$, whereas 
the transverse hydrostatic contribution scales as $
\delta\!\left(\partial_y \mathcal{H}\right)\sim \delta^{1-\alpha}
$. Balancing these two terms gives
$
\delta^\alpha\sim\delta^{1-\alpha}
$, hence
$
\alpha=\frac{1}{2}
$
and
$
\zeta=y t^{1/4}.
$

{T}he regularizing region width scales as $y\sim t^{-1/4}$. 
This distinguishes the saddle layer solution from both classical boundary layers \cite{vandyke1975perturbation} and self-similar solutions \cite{xue2020self}: its width is dynamically selected rather than imposed by a fixed geometrical length, while its amplitude remains tied to the logarithmic drainage profile.
We now restrict the analysis to the symmetry plane $x=0$, where $\partial_x \mathcal{H}=0$, to compare with the experimental ridge profile. The resulting ordinary differential equation reads
\begin{equation}
\mathcal{H}''\mathcal{H}^2+3\mathcal{H}\mathcal{H}'(\mathcal{H}'+\zeta)+\frac{3}{2}=0,
\label{eq:innerhydro}
\end{equation}
where primes denote derivatives with respect to $\zeta=yt^{1/4}$. When solved for $\zeta \geq 0$, the solution must satisfy the symmetry condition
$
\mathcal{H}'(0)=0,
$
and match the logarithmic behavior,
$
\mathcal{H}(\zeta\gg1)\sim H_d^* = \left(C-\log (\zeta t^{-1/4})\right)^{1/2}.
$
Thus the leading-order time dependence enters through the matching condition.
The large-$\zeta$ asymptotics of Eq.~\eqref{eq:innerhydro} recovers the  logarithmic form~\eqref{eq:outerlog}, allowing numerical matching at a single point in the overlap region. The matched solution reproduces the numerical profiles close to the saddle for the model without capillarity [Fig.~{\ref{fig:2}b}], thereby confirming the local balance mechanism. In particular, the numerical rescaled saddle thickness ${H}_s:={H}(0,0)$ collapses for different initial conditions and increases slowly with time [inset of Fig.~{\ref{fig:2}b}], consistent with the persistent sharpening of the ridge observed in the simulations. Although the physical thickness \(h=t^{-1/2}{\mathcal{H}}\) decrease{s}, the growth of \(\mathcal{H}\) shows that thinning is asymptotically slower there than far from the saddle{, signalling a local relative accumulation of fluid}.

\begin{figure}[t]
    \centering
    \includegraphics[width=\linewidth]{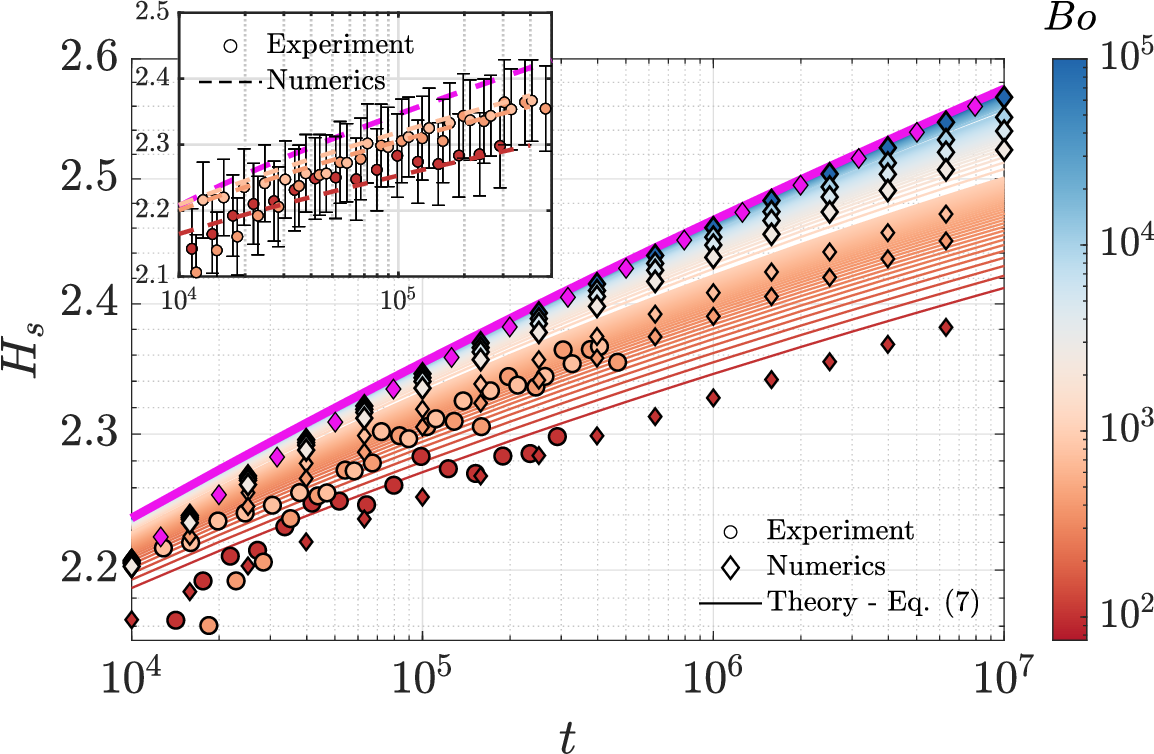}
    \caption{Regularization of the saddle singularity along $x=0$. Finite-capillarity case: rescaled saddle thickness $H_s=H(0,0)$ versus time. Diamonds are numerical solutions of $H$ (with $h_0=0.1$) for $Bo \in [10{2}, 10^5]${, from the bottom to the top: $Bo=102,408,725$ (same values as in the experiments), and $Bo=2500,5000,10^4,10^5$.} {{C}ircles are experimental measurements  with $Bo=10{2, 408},725$,}  colored lines are finite-capillarity saddle-layer predictions, and the magenta line shows the capillary-free limit. In the inset, {experiments are reported with their error bars, while the corresponding numerical simulations are shown as dashed lines, and the magenta dashed line is the numerical case without capillarity}.}
    \label{fig:3}
\end{figure}

At finite Bond number, numerical solutions deviate systematically from the capillary-free prediction [Fig.~\ref{fig:3}]. The deviation becomes stronger as the saddle layer narrows with time, indicating that capillary smoothing becomes comparable to the local hydrostatic correction in Eq.~\eqref{eq:localmodeldimless}.
Near the saddle, the leading transverse balance is between drainage, hydrostatic pressure and capillarity,
{$
\partial_y\left(\mathcal{H}^3 y\right)
\sim
\delta\,\partial_y\left(\mathcal{H}^3\partial_y \mathcal{H}\right)
\sim
{Bo}^{-1}{\delta}\,\partial_y\left(\mathcal{H}^3\partial_y^3 \mathcal{H}\right).
\label{eq:triplebalancephys}
$}
The saddle layer generates its own transverse scale $\ell_b=\hat\ell_b/L \sim \delta^{1/2} = t^{-1/4}$ and, therefore,
its own modified Bond number $Bo^*=\frac{\rho g \hat\ell_b^2}{\gamma}$. Capillarity enters in the dominant balance with drainage and hydrostatic pressure when $Bo^*=\mathcal{O}(1)$, i.e. under the distinguished scaling $Bo=Bo^*\,\delta^{-1}=Bo^* t^{1/2}$.
The dominant balances, using $\partial_y \sim \delta^{-\alpha}$, are
$
\delta^\alpha\sim\delta^{1-\alpha}\sim\delta^{2-3\alpha},
$
which give
$
\alpha=\frac{1}{2}.
$ 
Capillarity therefore enters the same $t^{-1/4}$ ridge width scaling without altering it, leading to the local equation
\begin{multline}
{\mathcal{H}}''{\mathcal{H}}^2-\frac{1}{Bo^*}\left(3{\mathcal{H}}{\mathcal{H}}'{\mathcal{H}}'''+{\mathcal{H}}^2{\mathcal{H}}^{\mathrm{IV}}\right)
\\ +3{\mathcal{H}}{\mathcal{H}}'({\mathcal{H}}'+\zeta)+\frac{3}{2}=0,
\label{eq:innercap}
\end{multline}
subject to the symmetry conditions 
${\mathcal{H}}'(0)={\mathcal{H}}'''(0)=0,$
together with matching to the logarithmic profile for both value and slope. 
The resulting solutions agree well with both numerical simulations and experiments [Fig.~\ref{fig:3}], without any fitting parameter, confirming that capillarity modifies the ridge profile but not the underlying ridge width scaling.
Residual small discrepancies in Fig.~\ref{fig:3} can be attributed, at early times, to the film not yet reaching the asymptotic $t^{-1/2}$ regime far from the saddle and the matching occurring in a region where the local metric is no longer well approximated as Cartesian, and finite Bond number effects in the drainage solution \cite{roy2002lubrication}.
{Furthermore, the comparatively larger experimental-numerical discrepancy observed for $Bo = 408$ likely results from the 3D printing resolution, whose magnitude relative to the geometric length scale $L$ is largest in this case }\cite{Note1}.
Eventually, a systematic expansion of the full model Eq. (1) leads to the same dominant balances and problems studied here, as shown in the Supplemental Material  \cite{Note1}.
{Additional measurements performed with a lower-viscosity silicone 
oil ($\nu=350 \pm 17 \,\mathrm{mm^2\,s^{-1}}$) show the
same redistribution mechanism \cite{Note1}.}

{However, real saddles do not always balance gravity exactly along the principal directions, and fabrication tolerances can introduce a small nonzero curvature. Its effect is quantifiable and leaves the mechanism intact, as follows. A generic saddle with nonzero mean curvature may be written locally as
$
\eta\simeq \frac{1}{2}(y^2- (1-\mathcal K_0) x^2)$ with $\mathcal K_0$ being twice the saddle mean curvature, here assumed to be small. Along $x=0$, Eq.~\eqref{eq:outerchar} becomes
$y\frac{\mathrm d H_d^2}{\mathrm dy}
+\frac{2}{3}\mathcal K_0 H_d^2+1=0
$, whose solution reads ($y>0$)
\begin{equation}
H_d=
\left( \frac{3}{2\mathcal K_0}
\left[
\left(\frac{y}{\exp[C]}\right)^{-2\mathcal K_0/3}-1
\right] \right)^{1/2} :=\tilde{H}_d, 
\label{eq:generic}
\end{equation}
where $C$ is set by global drainage, as before. 
For $\mathcal K_0>0$, $H_d$ diverges algebraically, whereas for $\mathcal K_0<0$
it approaches the value $\left(-3/(2\mathcal K_0)\right)^{1/2}$, with an infinite slope for $-3/2<\mathcal K_0<0$. 
However, drainage holds only up to the saddle layer, whose  $t^{-1/4}$ width scaling follows from Eq.~\eqref{eq:localmodeldimless}, unaffected by $\mathcal K_0$, and imposes small slopes and $H_d=\mathcal{O}(1)$.
Consequently, the power law can be converted into an exponential, namely $({y}/{\exp[C]})^{-2\mathcal K_0/3}=\exp[\frac{2\mathcal K_0}{3}( C-\log{y}) ] =\exp[\frac{2}{3} \mathcal{K}_0 H_d^{*2} ]$, 
and $\tilde{H}_d$ expanded for small $\mathcal{K}_0 H_d^{*2}$:\begin{equation}\tilde{H}_d = H_d^*
\left[
1+\frac{\mathcal K_0}{6}(H_d^*)^2
+O\!\left(\mathcal K_0^2(H_d^*)^4\right)
\right].\label{eq:expgen}\end{equation}
Hence, for saddles with comparable principal curvatures, a small nonzero mean curvature produces only a perturbative correction to the logarithmic profile and leaves the  regularization-layer width scaling unchanged. The ridge-formation mechanism therefore persists for a broad class of saddles with comparable principal curvatures.
}
{Because the drainage competition depends only on the local hyperbolic structure, it should arise generically on complex topographies, as numerically confirmed for another global substrate shape {\cite{Note1}}.}

{In summary, }{a} {smooth} saddle generates a {logarithmic ridge within a through-flow drainage state regularized by a dynamically selected region}. {Therefore, smooth saddles are not neutral elements on complex surfaces and act as organizing centers for coating flows alongside contact lines, substrate edges and defects. Restoring dimensions, the following scaling laws hold: }
{
\begin{equation}
\frac{\hat h_s}{\hat\ell_b}\sim
\frac{\hat\ell_b}{L}
\left[
C+\log\left(\frac{L}{\hat\ell_b}\right)
\right]^{1/2},
\,
\hat\ell_b\sim
L\left(\frac{\nu}{gL\hat t}\right)^{1/4},
\end{equation}
obtained by assuming $\zeta \sim 1$ at matching. Eliminating time yields a geometric relation between the instantaneous saddle thickness $\hat h_s$ and ridge width $\hat{\ell}_b$ that contains no explicit dependence on viscosity, gravity and drainage time. This relation may guide local geometric compensation before shell fabrication.
Within the Newtonian case considered here, the representative coating parameters for investment casting
$\nu=10^{-4}\,\mathrm{m^2\,s^{-1}}$,
$L=10\,\mathrm{mm}$ and $\hat t=60\,\mathrm{s}$ give a ridge approximately $0.6\,$ mm wide and a saddle thickness of order $10^2\,\mu$m, corresponding to an increase in maximum thickness of $\approx 100\%$ compared to the one developing on a cylindrical surface of radius equal to $L$ \cite{takagi2010flow}, but in a ridge ten times narrower. 
At geophysical scales, the geometric scaling may help infer topography and lava properties from preserved ridges in mountain-like passes \cite{garry2007morphology}. These examples motivate extensions to interactions among multiple geometrical features, evolving substrates \cite{carpy2024fingerprints}, complex and time-dependent rheology, as well as continuous injection and downstream evolution of propagating currents.
}

\begin{acknowledgments}

\textit{Acknowledgments --} S.D. acknowledges support from the Swiss National Science Foundation under grant no. 10001049. P.G.L. acknowledges H.A.~Stone, P.-T.~Brun, and G.~Lerisson for having raised this problem, separately, during the same visit to Princeton in 2021. P.G.L. also thanks F.~Giannetti for valuable suggestions on the matching procedure. F.G. thanks H.A. Stone for fruitful discussions. The authors are grateful to E. Jambon-Puillet for feedback on the manuscript. 

P.G.L. and F.G. conceived the project. P.G.L. supervised the research and coordinated the research group. P.G.L., S.D. and F.G. developed the theory.  P.G.L. and S.D. prepared and performed the numerical simulations. A.M. and P.G.L. designed and prepared the experiments. A.M. performed the experiments and postprocessed the data. P.G.L. prepared the draft. All authors contributed to the revision of the manuscript. 

\textit{Data availability --} The data that support the findings of this article are openly available \cite{zenodo_rep}

\end{acknowledgments}

\bibliography{biblio} 

\begin{thebibliography}{10}%
\makeatletter
\providecommand \@ifxundefined [1]{%
 \@ifx{#1\undefined}
}%
\providecommand \@ifnum [1]{%
 \ifnum #1\expandafter \@firstoftwo
 \else \expandafter \@secondoftwo
 \fi
}%
\providecommand \@ifx [1]{%
 \ifx #1\expandafter \@firstoftwo
 \else \expandafter \@secondoftwo
 \fi
}%
\providecommand \natexlab [1]{#1}%
\providecommand \enquote  [1]{``#1''}%
\providecommand \bibnamefont  [1]{#1}%
\providecommand \bibfnamefont [1]{#1}%
\providecommand \citenamefont [1]{#1}%
\providecommand \href@noop [0]{\@secondoftwo}%
\providecommand \href [0]{\begingroup \@sanitize@url \@href}%
\providecommand \@href[1]{\@@startlink{#1}\@@href}%
\providecommand \@@href[1]{\endgroup#1\@@endlink}%
\providecommand \@sanitize@url [0]{\catcode `\\12\catcode `\$12\catcode
  `\&12\catcode `\#12\catcode `\^12\catcode `\_12\catcode `\%12\relax}%
\providecommand \@@startlink[1]{}%
\providecommand \@@endlink[0]{}%
\providecommand \url  [0]{\begingroup\@sanitize@url \@url }%
\providecommand \@url [1]{\endgroup\@href {#1}{\urlprefix }}%
\providecommand \urlprefix  [0]{URL }%
\providecommand \Eprint [0]{\href }%
\providecommand \doibase [0]{http://dx.doi.org/}%
\providecommand \selectlanguage [0]{\@gobble}%
\providecommand \bibinfo  [0]{\@secondoftwo}%
\providecommand \bibfield  [0]{\@secondoftwo}%
\providecommand \translation [1]{[#1]}%
\providecommand \BibitemOpen [0]{}%
\providecommand \bibitemStop [0]{}%
\providecommand \bibitemNoStop [0]{.\EOS\space}%
\providecommand \EOS [0]{\spacefactor3000\relax}%
\providecommand \BibitemShut  [1]{\csname bibitem#1\endcsname}%
\let\auto@bib@innerbib\@empty
\bibitem [{\citenamefont {Wang}\ \emph {et~al.}(2026)\citenamefont {Wang},
  \citenamefont {Chen}, \citenamefont {Constante-Amores}, \citenamefont
  {Gorse}, \citenamefont {Castrej\'on-Pita},\ and\ \citenamefont
  {Castrej\'on-Pita}}]{wang2026}%
  \BibitemOpen
  \bibfield  {author} {\bibinfo {author} {\bibfnamefont {J.}~\bibnamefont
  {Wang}}, \bibinfo {author} {\bibfnamefont {Q.}~\bibnamefont {Chen}}, \bibinfo
  {author} {\bibfnamefont {C.~R.}\ \bibnamefont {Constante-Amores}}, \bibinfo
  {author} {\bibfnamefont {D.}~\bibnamefont {Gorse}}, \bibinfo {author}
  {\bibfnamefont {A.~A.}\ \bibnamefont {Castrej\'on-Pita}}, \ and\ \bibinfo
  {author} {\bibfnamefont {J.~R.}\ \bibnamefont {Castrej\'on-Pita}},\
  }\href@noop {} {\bibfield  {journal} {\bibinfo  {journal} {Phys. Rev. X}\
  }\textbf {\bibinfo {volume} {16}},\ \bibinfo {pages} {031021} (\bibinfo
  {year} {2026})}\BibitemShut {NoStop}%
\bibitem [{\citenamefont {Irgens}\ \emph {et~al.}(2019)\citenamefont {Irgens},
  \citenamefont {Irgens},\ and\ \citenamefont {Baumann}}]{irgens2019tensor}%
  \BibitemOpen
  \bibfield  {author} {\bibinfo {author} {\bibfnamefont {F.}~\bibnamefont
  {Irgens}}, \bibinfo {author} {\bibfnamefont {B.}~\bibnamefont {Irgens}}, \
  and\ \bibinfo {author} {\bibfnamefont {D.}~\bibnamefont {Baumann}},\
  }\href@noop {} {\emph {\bibinfo {title} {Tensor analysis}}}\ (\bibinfo
  {publisher} {Springer},\ \bibinfo {year} {2019})\BibitemShut {NoStop}%
\bibitem [{\citenamefont {Roy}\ \emph {et~al.}(2002)\citenamefont {Roy},
  \citenamefont {Roberts},\ and\ \citenamefont {Simpson}}]{roy2002lubrication}%
  \BibitemOpen
  \bibfield  {author} {\bibinfo {author} {\bibfnamefont {R.~V.}\ \bibnamefont
  {Roy}}, \bibinfo {author} {\bibfnamefont {A.}~\bibnamefont {Roberts}}, \ and\
  \bibinfo {author} {\bibfnamefont {M.}~\bibnamefont {Simpson}},\ }\href@noop
  {} {\bibfield  {journal} {\bibinfo  {journal} {J. Fluid Mech.}\ }\textbf
  {\bibinfo {volume} {454}},\ \bibinfo {pages} {235} (\bibinfo {year}
  {2002})}\BibitemShut {NoStop}%
\bibitem [{\citenamefont {Thiffeault}\ and\ \citenamefont
  {Kamhawi}(2006)}]{thiffeault2006transport}%
  \BibitemOpen
  \bibfield  {author} {\bibinfo {author} {\bibfnamefont {J.-L.}\ \bibnamefont
  {Thiffeault}}\ and\ \bibinfo {author} {\bibfnamefont {K.}~\bibnamefont
  {Kamhawi}},\ }\href@noop {} {\bibfield  {journal} {\bibinfo  {journal} {arXiv
  preprint nlin/0607075}\ } (\bibinfo {year} {2006})}\BibitemShut {NoStop}%
\bibitem [{\citenamefont {Wray}\ \emph {et~al.}(2017)\citenamefont {Wray},
  \citenamefont {Papageorgiou},\ and\ \citenamefont {Matar}}]{wray2017reduced}%
  \BibitemOpen
  \bibfield  {author} {\bibinfo {author} {\bibfnamefont {A.~W.}\ \bibnamefont
  {Wray}}, \bibinfo {author} {\bibfnamefont {D.~T.}\ \bibnamefont
  {Papageorgiou}}, \ and\ \bibinfo {author} {\bibfnamefont {O.~K.}\
  \bibnamefont {Matar}},\ }\href@noop {} {\bibfield  {journal} {\bibinfo
  {journal} {SIAM J. Appl. Math.}\ }\textbf {\bibinfo {volume} {77}},\ \bibinfo
  {pages} {881} (\bibinfo {year} {2017})}\BibitemShut {NoStop}%
\bibitem [{\citenamefont {Shepherd}\ \emph {et~al.}(2024)\citenamefont
  {Shepherd}, \citenamefont {Boujo},\ and\ \citenamefont
  {Sellier}}]{shepherd2024general}%
  \BibitemOpen
  \bibfield  {author} {\bibinfo {author} {\bibfnamefont {R.~G.}\ \bibnamefont
  {Shepherd}}, \bibinfo {author} {\bibfnamefont {E.}~\bibnamefont {Boujo}}, \
  and\ \bibinfo {author} {\bibfnamefont {M.}~\bibnamefont {Sellier}},\
  }\href@noop {} {\bibfield  {journal} {\bibinfo  {journal} {J. Fluid Mech.}\
  }\textbf {\bibinfo {volume} {998}},\ \bibinfo {pages} {A32} (\bibinfo {year}
  {2024})}\BibitemShut {NoStop}%
\bibitem [{\citenamefont {Howell}(2003)}]{howell2003surface}%
  \BibitemOpen
  \bibfield  {author} {\bibinfo {author} {\bibfnamefont {P.~D.}\ \bibnamefont
  {Howell}},\ }\href@noop {} {\bibfield  {journal} {\bibinfo  {journal} {J.
  Eng. Math.}\ }\textbf {\bibinfo {volume} {45}},\ \bibinfo {pages} {283}
  (\bibinfo {year} {2003})}\BibitemShut {NoStop}%
\bibitem [{\citenamefont {Roberts}\ and\ \citenamefont
  {Li}(2006)}]{roberts2006accurate}%
  \BibitemOpen
  \bibfield  {author} {\bibinfo {author} {\bibfnamefont {A.~J.}\ \bibnamefont
  {Roberts}}\ and\ \bibinfo {author} {\bibfnamefont {Z.}~\bibnamefont {Li}},\
  }\href@noop {} {\bibfield  {journal} {\bibinfo  {journal} {J. Fluid Mech.}\
  }\textbf {\bibinfo {volume} {553}},\ \bibinfo {pages} {33} (\bibinfo {year}
  {2006})}\BibitemShut {NoStop}%
\bibitem [{\citenamefont {Ledda}\ \emph {et~al.}(2022)\citenamefont {Ledda},
  \citenamefont {Pezzulla}, \citenamefont {Jambon-Puillet}, \citenamefont
  {Brun},\ and\ \citenamefont {Gallaire}}]{ledda2022gravity}%
  \BibitemOpen
  \bibfield  {author} {\bibinfo {author} {\bibfnamefont {P.~G.}\ \bibnamefont
  {Ledda}}, \bibinfo {author} {\bibfnamefont {M.}~\bibnamefont {Pezzulla}},
  \bibinfo {author} {\bibfnamefont {E.}~\bibnamefont {Jambon-Puillet}},
  \bibinfo {author} {\bibfnamefont {P.-T.}\ \bibnamefont {Brun}}, \ and\
  \bibinfo {author} {\bibfnamefont {F.}~\bibnamefont {Gallaire}},\ }\href@noop
  {} {\bibfield  {journal} {\bibinfo  {journal} {J. Fluid Mech.}\ }\textbf
  {\bibinfo {volume} {949}},\ \bibinfo {pages} {A38} (\bibinfo {year}
  {2022})}\BibitemShut {NoStop}%
\bibitem [{\citenamefont {Bers}\ \emph {et~al.}(1964)\citenamefont {Bers},
  \citenamefont {John},\ and\ \citenamefont {Schechter}}]{bers1964partial}%
  \BibitemOpen
  \bibfield  {author} {\bibinfo {author} {\bibfnamefont {L.}~\bibnamefont
  {Bers}}, \bibinfo {author} {\bibfnamefont {F.}~\bibnamefont {John}}, \ and\
  \bibinfo {author} {\bibfnamefont {M.}~\bibnamefont {Schechter}},\ }\href@noop
  {} {\emph {\bibinfo {title} {Partial differential equations}}}\ (\bibinfo
  {publisher} {American Mathematical Soc.},\ \bibinfo {year}
  {1964})\BibitemShut {NoStop}%
\end{thebibliography}%


\begin{thebibliography}{35}%
\makeatletter
\providecommand \@ifxundefined [1]{%
 \@ifx{#1\undefined}
}%
\providecommand \@ifnum [1]{%
 \ifnum #1\expandafter \@firstoftwo
 \else \expandafter \@secondoftwo
 \fi
}%
\providecommand \@ifx [1]{%
 \ifx #1\expandafter \@firstoftwo
 \else \expandafter \@secondoftwo
 \fi
}%
\providecommand \natexlab [1]{#1}%
\providecommand \enquote  [1]{``#1''}%
\providecommand \bibnamefont  [1]{#1}%
\providecommand \bibfnamefont [1]{#1}%
\providecommand \citenamefont [1]{#1}%
\providecommand \href@noop [0]{\@secondoftwo}%
\providecommand \href [0]{\begingroup \@sanitize@url \@href}%
\providecommand \@href[1]{\@@startlink{#1}\@@href}%
\providecommand \@@href[1]{\endgroup#1\@@endlink}%
\providecommand \@sanitize@url [0]{\catcode `\\12\catcode `\$12\catcode
  `\&12\catcode `\#12\catcode `\^12\catcode `\_12\catcode `\%12\relax}%
\providecommand \@@startlink[1]{}%
\providecommand \@@endlink[0]{}%
\providecommand \url  [0]{\begingroup\@sanitize@url \@url }%
\providecommand \@url [1]{\endgroup\@href {#1}{\urlprefix }}%
\providecommand \urlprefix  [0]{URL }%
\providecommand \Eprint [0]{\href }%
\providecommand \doibase [0]{http://dx.doi.org/}%
\providecommand \selectlanguage [0]{\@gobble}%
\providecommand \bibinfo  [0]{\@secondoftwo}%
\providecommand \bibfield  [0]{\@secondoftwo}%
\providecommand \translation [1]{[#1]}%
\providecommand \BibitemOpen [0]{}%
\providecommand \bibitemStop [0]{}%
\providecommand \bibitemNoStop [0]{.\EOS\space}%
\providecommand \EOS [0]{\spacefactor3000\relax}%
\providecommand \BibitemShut  [1]{\csname bibitem#1\endcsname}%
\let\auto@bib@innerbib\@empty
\bibitem [{\citenamefont {Scriven}(1988)}]{scriven1988physics}%
  \BibitemOpen
  \bibfield  {author} {\bibinfo {author} {\bibfnamefont {L.}~\bibnamefont
  {Scriven}},\ }\href@noop {} {\bibfield  {journal} {\bibinfo  {journal} {MRS
  Online Proc. Libr.}\ }\textbf {\bibinfo {volume} {121}},\ \bibinfo {pages}
  {717} (\bibinfo {year} {1988})}\BibitemShut {NoStop}%
\bibitem [{\citenamefont {Weinstein}\ and\ \citenamefont
  {Ruschak}(2004)}]{weinstein2004coating}%
  \BibitemOpen
  \bibfield  {author} {\bibinfo {author} {\bibfnamefont {S.~J.}\ \bibnamefont
  {Weinstein}}\ and\ \bibinfo {author} {\bibfnamefont {K.~J.}\ \bibnamefont
  {Ruschak}},\ }\href@noop {} {\bibfield  {journal} {\bibinfo  {journal} {Annu.
  Rev. Fluid Mech.}\ }\textbf {\bibinfo {volume} {36}},\ \bibinfo {pages} {29}
  (\bibinfo {year} {2004})}\BibitemShut {NoStop}%
\bibitem [{\citenamefont {Neufeld}\ \emph {et~al.}(2011)\citenamefont
  {Neufeld}, \citenamefont {Vella}, \citenamefont {Huppert},\ and\
  \citenamefont {Lister}}]{neufeld2011leakage}%
  \BibitemOpen
  \bibfield  {author} {\bibinfo {author} {\bibfnamefont {J.~A.}\ \bibnamefont
  {Neufeld}}, \bibinfo {author} {\bibfnamefont {D.}~\bibnamefont {Vella}},
  \bibinfo {author} {\bibfnamefont {H.~E.}\ \bibnamefont {Huppert}}, \ and\
  \bibinfo {author} {\bibfnamefont {J.~R.}\ \bibnamefont {Lister}},\
  }\href@noop {} {\bibfield  {journal} {\bibinfo  {journal} {J. Fluid Mech.}\
  }\textbf {\bibinfo {volume} {666}},\ \bibinfo {pages} {391} (\bibinfo {year}
  {2011})}\BibitemShut {NoStop}%
\bibitem [{\citenamefont {Huppert}(1986)}]{huppert1986intrusion}%
  \BibitemOpen
  \bibfield  {author} {\bibinfo {author} {\bibfnamefont {H.~E.}\ \bibnamefont
  {Huppert}},\ }\href@noop {} {\bibfield  {journal} {\bibinfo  {journal} {J.
  Fluid Mech.}\ }\textbf {\bibinfo {volume} {173}},\ \bibinfo {pages} {557}
  (\bibinfo {year} {1986})}\BibitemShut {NoStop}%
\bibitem [{\citenamefont {Meakin}\ and\ \citenamefont
  {Jamtveit}(2010)}]{meakin2010geological}%
  \BibitemOpen
  \bibfield  {author} {\bibinfo {author} {\bibfnamefont {P.}~\bibnamefont
  {Meakin}}\ and\ \bibinfo {author} {\bibfnamefont {B.}~\bibnamefont
  {Jamtveit}},\ }\href@noop {} {\bibfield  {journal} {\bibinfo  {journal}
  {Proc. R. Soc. A: Math. Phys. Eng. Sci.}\ }\textbf {\bibinfo {volume}
  {466}},\ \bibinfo {pages} {659} (\bibinfo {year} {2010})}\BibitemShut
  {NoStop}%
\bibitem [{\citenamefont {Bertagni}\ and\ \citenamefont
  {Camporeale}(2021)}]{bertagni2021hydrodynamic}%
  \BibitemOpen
  \bibfield  {author} {\bibinfo {author} {\bibfnamefont {M.}~\bibnamefont
  {Bertagni}}\ and\ \bibinfo {author} {\bibfnamefont {C.}~\bibnamefont
  {Camporeale}},\ }\href@noop {} {\bibfield  {journal} {\bibinfo  {journal} {J.
  Fluid Mech.}\ }\textbf {\bibinfo {volume} {913}},\ \bibinfo {pages} {A34}
  (\bibinfo {year} {2021})}\BibitemShut {NoStop}%
\bibitem [{\citenamefont {Dong}\ \emph {et~al.}(2024)\citenamefont {Dong},
  \citenamefont {Wang}, \citenamefont {Cui}, \citenamefont {Wang},
  \citenamefont {Wang}, \citenamefont {Kang},\ and\ \citenamefont
  {Jiang}}]{dong2024shell}%
  \BibitemOpen
  \bibfield  {author} {\bibinfo {author} {\bibfnamefont {R.-z.}\ \bibnamefont
  {Dong}}, \bibinfo {author} {\bibfnamefont {W.-h.}\ \bibnamefont {Wang}},
  \bibinfo {author} {\bibfnamefont {K.}~\bibnamefont {Cui}}, \bibinfo {author}
  {\bibfnamefont {Y.-b.}\ \bibnamefont {Wang}}, \bibinfo {author}
  {\bibfnamefont {Z.-c.}\ \bibnamefont {Wang}}, \bibinfo {author}
  {\bibfnamefont {K.-y.}\ \bibnamefont {Kang}}, \ and\ \bibinfo {author}
  {\bibfnamefont {R.-s.}\ \bibnamefont {Jiang}},\ }\href@noop {} {\bibfield
  {journal} {\bibinfo  {journal} {Journal of Manufacturing Processes}\ }\textbf
  {\bibinfo {volume} {131}},\ \bibinfo {pages} {507} (\bibinfo {year}
  {2024})}\BibitemShut {NoStop}%
\bibitem [{\citenamefont {Saville}\ \emph {et~al.}(2022)\citenamefont
  {Saville}, \citenamefont {Hinton},\ and\ \citenamefont
  {Huppert}}]{saville2022safe}%
  \BibitemOpen
  \bibfield  {author} {\bibinfo {author} {\bibfnamefont {J.~M.}\ \bibnamefont
  {Saville}}, \bibinfo {author} {\bibfnamefont {E.~M.}\ \bibnamefont {Hinton}},
  \ and\ \bibinfo {author} {\bibfnamefont {H.~E.}\ \bibnamefont {Huppert}},\
  }\href@noop {} {\bibfield  {journal} {\bibinfo  {journal} {J. Geophys. Res.
  Solid Earth}\ }\textbf {\bibinfo {volume} {127}},\ \bibinfo {pages}
  {e2022JB024167} (\bibinfo {year} {2022})}\BibitemShut {NoStop}%
\bibitem [{\citenamefont {Younger}\ \emph {et~al.}(2019)\citenamefont
  {Younger}, \citenamefont {Valentine},\ and\ \citenamefont
  {Gregg}}]{younger2019marcath}%
  \BibitemOpen
  \bibfield  {author} {\bibinfo {author} {\bibfnamefont {Z.~P.}\ \bibnamefont
  {Younger}}, \bibinfo {author} {\bibfnamefont {G.~A.}\ \bibnamefont
  {Valentine}}, \ and\ \bibinfo {author} {\bibfnamefont {T.~K.~P.}\
  \bibnamefont {Gregg}},\ }\href@noop {} {\bibfield  {journal} {\bibinfo
  {journal} {Bull. Volcanol.}\ }\textbf {\bibinfo {volume} {81}},\ \bibinfo
  {pages} {50} (\bibinfo {year} {2019})}\BibitemShut {NoStop}%
\bibitem [{\citenamefont {Oron}\ \emph {et~al.}(1997)\citenamefont {Oron},
  \citenamefont {Davis},\ and\ \citenamefont {Bankoff}}]{oron1997long}%
  \BibitemOpen
  \bibfield  {author} {\bibinfo {author} {\bibfnamefont {A.}~\bibnamefont
  {Oron}}, \bibinfo {author} {\bibfnamefont {S.~H.}\ \bibnamefont {Davis}}, \
  and\ \bibinfo {author} {\bibfnamefont {S.~G.}\ \bibnamefont {Bankoff}},\
  }\href@noop {} {\bibfield  {journal} {\bibinfo  {journal} {Rev. Mod. Phys.}\
  }\textbf {\bibinfo {volume} {69}},\ \bibinfo {pages} {931} (\bibinfo {year}
  {1997})}\BibitemShut {NoStop}%
\bibitem [{\citenamefont {Roy}\ \emph {et~al.}(2002)\citenamefont {Roy},
  \citenamefont {Roberts},\ and\ \citenamefont {Simpson}}]{roy2002lubrication}%
  \BibitemOpen
  \bibfield  {author} {\bibinfo {author} {\bibfnamefont {R.~V.}\ \bibnamefont
  {Roy}}, \bibinfo {author} {\bibfnamefont {A.}~\bibnamefont {Roberts}}, \ and\
  \bibinfo {author} {\bibfnamefont {M.}~\bibnamefont {Simpson}},\ }\href@noop
  {} {\bibfield  {journal} {\bibinfo  {journal} {J. Fluid Mech.}\ }\textbf
  {\bibinfo {volume} {454}},\ \bibinfo {pages} {235} (\bibinfo {year}
  {2002})}\BibitemShut {NoStop}%
\bibitem [{\citenamefont {Lee}\ \emph {et~al.}(2016)\citenamefont {Lee},
  \citenamefont {Brun}, \citenamefont {Marthelot}, \citenamefont {Balestra},
  \citenamefont {Gallaire},\ and\ \citenamefont {Reis}}]{lee2016fabrication}%
  \BibitemOpen
  \bibfield  {author} {\bibinfo {author} {\bibfnamefont {A.}~\bibnamefont
  {Lee}}, \bibinfo {author} {\bibfnamefont {P.-T.}\ \bibnamefont {Brun}},
  \bibinfo {author} {\bibfnamefont {J.}~\bibnamefont {Marthelot}}, \bibinfo
  {author} {\bibfnamefont {G.}~\bibnamefont {Balestra}}, \bibinfo {author}
  {\bibfnamefont {F.}~\bibnamefont {Gallaire}}, \ and\ \bibinfo {author}
  {\bibfnamefont {P.~M.}\ \bibnamefont {Reis}},\ }\href@noop {} {\bibfield
  {journal} {\bibinfo  {journal} {Nat. Comm.}\ }\textbf {\bibinfo {volume}
  {7}},\ \bibinfo {pages} {11155} (\bibinfo {year} {2016})}\BibitemShut
  {NoStop}%
\bibitem [{\citenamefont {Balestra}\ \emph {et~al.}(2019)\citenamefont
  {Balestra}, \citenamefont {Badaoui}, \citenamefont {Ducimeti{\`e}re},\ and\
  \citenamefont {Gallaire}}]{balestra2019fingering}%
  \BibitemOpen
  \bibfield  {author} {\bibinfo {author} {\bibfnamefont {G.}~\bibnamefont
  {Balestra}}, \bibinfo {author} {\bibfnamefont {M.}~\bibnamefont {Badaoui}},
  \bibinfo {author} {\bibfnamefont {Y.-M.}\ \bibnamefont {Ducimeti{\`e}re}}, \
  and\ \bibinfo {author} {\bibfnamefont {F.}~\bibnamefont {Gallaire}},\
  }\href@noop {} {\bibfield  {journal} {\bibinfo  {journal} {J. Fluid Mech.}\
  }\textbf {\bibinfo {volume} {868}},\ \bibinfo {pages} {726} (\bibinfo {year}
  {2019})}\BibitemShut {NoStop}%
\bibitem [{\citenamefont {Ledda}\ \emph {et~al.}(2022)\citenamefont {Ledda},
  \citenamefont {Pezzulla}, \citenamefont {Jambon-Puillet}, \citenamefont
  {Brun},\ and\ \citenamefont {Gallaire}}]{ledda2022gravity}%
  \BibitemOpen
  \bibfield  {author} {\bibinfo {author} {\bibfnamefont {P.~G.}\ \bibnamefont
  {Ledda}}, \bibinfo {author} {\bibfnamefont {M.}~\bibnamefont {Pezzulla}},
  \bibinfo {author} {\bibfnamefont {E.}~\bibnamefont {Jambon-Puillet}},
  \bibinfo {author} {\bibfnamefont {P.-T.}\ \bibnamefont {Brun}}, \ and\
  \bibinfo {author} {\bibfnamefont {F.}~\bibnamefont {Gallaire}},\ }\href@noop
  {} {\bibfield  {journal} {\bibinfo  {journal} {J. Fluid Mech.}\ }\textbf
  {\bibinfo {volume} {949}},\ \bibinfo {pages} {A38} (\bibinfo {year}
  {2022})}\BibitemShut {NoStop}%
\bibitem [{\citenamefont {Jeffreys}(1930)}]{jeffreys1930draining}%
  \BibitemOpen
  \bibfield  {author} {\bibinfo {author} {\bibfnamefont {H.}~\bibnamefont
  {Jeffreys}},\ }\bibfield  {booktitle} {\emph {\bibinfo {booktitle} {Math.
  Proc. Camb. Phil. Soc.}},\ }\href@noop {} {\ \textbf {\bibinfo {volume}
  {26}},\ \bibinfo {pages} {204} (\bibinfo {year} {1930})}\BibitemShut
  {NoStop}%
\bibitem [{\citenamefont {Huppert}(1982)}]{huppert1982flow}%
  \BibitemOpen
  \bibfield  {author} {\bibinfo {author} {\bibfnamefont {H.~E.}\ \bibnamefont
  {Huppert}},\ }\href@noop {} {\bibfield  {journal} {\bibinfo  {journal}
  {Nature}\ }\textbf {\bibinfo {volume} {300}},\ \bibinfo {pages} {427}
  (\bibinfo {year} {1982})}\BibitemShut {NoStop}%
\bibitem [{\citenamefont {Lister}(1992)}]{lister1992viscous}%
  \BibitemOpen
  \bibfield  {author} {\bibinfo {author} {\bibfnamefont {J.~R.}\ \bibnamefont
  {Lister}},\ }\href@noop {} {\bibfield  {journal} {\bibinfo  {journal} {J.
  Fluid Mech.}\ }\textbf {\bibinfo {volume} {242}},\ \bibinfo {pages} {631}
  (\bibinfo {year} {1992})}\BibitemShut {NoStop}%
\bibitem [{\citenamefont {Xue}\ and\ \citenamefont
  {Stone}(2020)}]{xue2020self}%
  \BibitemOpen
  \bibfield  {author} {\bibinfo {author} {\bibfnamefont {N.}~\bibnamefont
  {Xue}}\ and\ \bibinfo {author} {\bibfnamefont {H.~A.}\ \bibnamefont
  {Stone}},\ }\href@noop {} {\bibfield  {journal} {\bibinfo  {journal} {Phys.
  Rev. Lett.}\ }\textbf {\bibinfo {volume} {125}},\ \bibinfo {pages} {064502}
  (\bibinfo {year} {2020})}\BibitemShut {NoStop}%
\bibitem [{\citenamefont {Takagi}\ and\ \citenamefont
  {Huppert}(2010)}]{takagi2010flow}%
  \BibitemOpen
  \bibfield  {author} {\bibinfo {author} {\bibfnamefont {D.}~\bibnamefont
  {Takagi}}\ and\ \bibinfo {author} {\bibfnamefont {H.~E.}\ \bibnamefont
  {Huppert}},\ }\href@noop {} {\bibfield  {journal} {\bibinfo  {journal} {J.
  Fluid Mech.}\ }\textbf {\bibinfo {volume} {647}},\ \bibinfo {pages} {221}
  (\bibinfo {year} {2010})}\BibitemShut {NoStop}%
\bibitem [{\citenamefont {Xue}\ and\ \citenamefont
  {Stone}(2021)}]{xue2021draining}%
  \BibitemOpen
  \bibfield  {author} {\bibinfo {author} {\bibfnamefont {N.}~\bibnamefont
  {Xue}}\ and\ \bibinfo {author} {\bibfnamefont {H.~A.}\ \bibnamefont
  {Stone}},\ }\href@noop {} {\bibfield  {journal} {\bibinfo  {journal} {Phys.
  Rev. Fluids}\ }\textbf {\bibinfo {volume} {6}},\ \bibinfo {pages} {043801}
  (\bibinfo {year} {2021})}\BibitemShut {NoStop}%
\bibitem [{\citenamefont {Perrin}\ \emph {et~al.}(2016)\citenamefont {Perrin},
  \citenamefont {Lhermerout}, \citenamefont {Davitt}, \citenamefont {Rolley},\
  and\ \citenamefont {Andreotti}}]{perrin2016defects}%
  \BibitemOpen
  \bibfield  {author} {\bibinfo {author} {\bibfnamefont {H.}~\bibnamefont
  {Perrin}}, \bibinfo {author} {\bibfnamefont {R.}~\bibnamefont {Lhermerout}},
  \bibinfo {author} {\bibfnamefont {K.}~\bibnamefont {Davitt}}, \bibinfo
  {author} {\bibfnamefont {E.}~\bibnamefont {Rolley}}, \ and\ \bibinfo {author}
  {\bibfnamefont {B.}~\bibnamefont {Andreotti}},\ }\href@noop {} {\bibfield
  {journal} {\bibinfo  {journal} {Phys. Rev. Lett.}\ }\textbf {\bibinfo
  {volume} {116}},\ \bibinfo {pages} {184502} (\bibinfo {year}
  {2016})}\BibitemShut {NoStop}%
\bibitem [{\citenamefont {He}\ and\ \citenamefont
  {Nagel}(2019)}]{he2019characteristic}%
  \BibitemOpen
  \bibfield  {author} {\bibinfo {author} {\bibfnamefont {M.}~\bibnamefont
  {He}}\ and\ \bibinfo {author} {\bibfnamefont {S.~R.}\ \bibnamefont {Nagel}},\
  }\href@noop {} {\bibfield  {journal} {\bibinfo  {journal} {Phys. Rev. Lett.}\
  }\textbf {\bibinfo {volume} {122}},\ \bibinfo {pages} {018001} (\bibinfo
  {year} {2019})}\BibitemShut {NoStop}%
\bibitem [{\citenamefont {Wang}\ \emph {et~al.}(2026)\citenamefont {Wang},
  \citenamefont {Chen}, \citenamefont {Constante-Amores}, \citenamefont
  {Gorse}, \citenamefont {Castrej\'on-Pita},\ and\ \citenamefont
  {Castrej\'on-Pita}}]{wang2026}%
  \BibitemOpen
  \bibfield  {author} {\bibinfo {author} {\bibfnamefont {J.}~\bibnamefont
  {Wang}}, \bibinfo {author} {\bibfnamefont {Q.}~\bibnamefont {Chen}}, \bibinfo
  {author} {\bibfnamefont {C.~R.}\ \bibnamefont {Constante-Amores}}, \bibinfo
  {author} {\bibfnamefont {D.}~\bibnamefont {Gorse}}, \bibinfo {author}
  {\bibfnamefont {A.~A.}\ \bibnamefont {Castrej\'on-Pita}}, \ and\ \bibinfo
  {author} {\bibfnamefont {J.~R.}\ \bibnamefont {Castrej\'on-Pita}},\
  }\href@noop {} {\bibfield  {journal} {\bibinfo  {journal} {Phys. Rev. X}\
  }\textbf {\bibinfo {volume} {16}},\ \bibinfo {pages} {031021} (\bibinfo
  {year} {2026})}\BibitemShut {NoStop}%
\bibitem [{Note1()}]{Note1}%
  \BibitemOpen
  \bibinfo {note} {See Supplemental Material for further detail, which includes
  Refs. \cite
  {ledda2022gravity,bers1964partial,roy2002lubrication,howell2003surface,roberts2006accurate,thiffeault2006transport,wray2017reduced,shepherd2024general,irgens2019tensor}}\BibitemShut
  {NoStop}%
\bibitem [{\citenamefont {Van~Dyke}(1975)}]{vandyke1975perturbation}%
  \BibitemOpen
  \bibfield  {author} {\bibinfo {author} {\bibfnamefont {M.}~\bibnamefont
  {Van~Dyke}},\ }\href@noop {} {\emph {\bibinfo {title} {Perturbation Methods
  in Fluid Mechanics}}}\ (\bibinfo  {publisher} {Parabolic Press},\ \bibinfo
  {address} {Stanford, Calif.},\ \bibinfo {year} {1975})\BibitemShut {NoStop}%
\bibitem [{\citenamefont {Garry}\ \emph {et~al.}(2007)\citenamefont {Garry},
  \citenamefont {Zimbelman},\ and\ \citenamefont
  {Gregg}}]{garry2007morphology}%
  \BibitemOpen
  \bibfield  {author} {\bibinfo {author} {\bibfnamefont {W.~B.}\ \bibnamefont
  {Garry}}, \bibinfo {author} {\bibfnamefont {J.~R.}\ \bibnamefont
  {Zimbelman}}, \ and\ \bibinfo {author} {\bibfnamefont {T.~K.~P.}\
  \bibnamefont {Gregg}},\ }\href@noop {} {\bibfield  {journal} {\bibinfo
  {journal} {J. Geophys. Res. Planets}\ }\textbf {\bibinfo {volume} {112}},\
  \bibinfo {pages} {E8} (\bibinfo {year} {2007})}\BibitemShut {NoStop}%
\bibitem [{\citenamefont {Carpy}\ \emph {et~al.}(2024)\citenamefont {Carpy},
  \citenamefont {Berhanu}, \citenamefont {Chaigne},\ and\ \citenamefont
  {Courrech Du~Pont}}]{carpy2024fingerprints}%
  \BibitemOpen
  \bibfield  {author} {\bibinfo {author} {\bibfnamefont {S.}~\bibnamefont
  {Carpy}}, \bibinfo {author} {\bibfnamefont {M.}~\bibnamefont {Berhanu}},
  \bibinfo {author} {\bibfnamefont {M.}~\bibnamefont {Chaigne}}, \ and\
  \bibinfo {author} {\bibfnamefont {S.}~\bibnamefont {Courrech Du~Pont}},\
  }\href@noop {} {\bibfield  {journal} {\bibinfo  {journal} {Comptes Rendus.
  Physique}\ }\textbf {\bibinfo {volume} {25}},\ \bibinfo {pages} {95}
  (\bibinfo {year} {2024})}\BibitemShut {NoStop}%
\bibitem [{\citenamefont {Djambov}\ \emph {et~al.}(2026)\citenamefont
  {Djambov}, \citenamefont {Marcotte}, \citenamefont {Gallaire},\ and\
  \citenamefont {Ledda}}]{zenodo_rep}%
  \BibitemOpen
  \bibfield  {author} {\bibinfo {author} {\bibfnamefont {S.}~\bibnamefont
  {Djambov}}, \bibinfo {author} {\bibfnamefont {A.}~\bibnamefont {Marcotte}},
  \bibinfo {author} {\bibfnamefont {F.}~\bibnamefont {Gallaire}}, \ and\
  \bibinfo {author} {\bibfnamefont {P.~G.}\ \bibnamefont {Ledda}},\ }\href@noop
  {} {\bibfield  {journal} {\bibinfo  {journal} {Zenodo}\ } (\bibinfo {year}
  {2026})},\ \bibinfo {note} {\textit{to be published upon
  acceptance}}\BibitemShut {NoStop}%
\bibitem [{\citenamefont {Bers}\ \emph {et~al.}(1964)\citenamefont {Bers},
  \citenamefont {John},\ and\ \citenamefont {Schechter}}]{bers1964partial}%
  \BibitemOpen
  \bibfield  {author} {\bibinfo {author} {\bibfnamefont {L.}~\bibnamefont
  {Bers}}, \bibinfo {author} {\bibfnamefont {F.}~\bibnamefont {John}}, \ and\
  \bibinfo {author} {\bibfnamefont {M.}~\bibnamefont {Schechter}},\ }\href@noop
  {} {\emph {\bibinfo {title} {Partial differential equations}}}\ (\bibinfo
  {publisher} {American Mathematical Soc.},\ \bibinfo {year}
  {1964})\BibitemShut {NoStop}%
\bibitem [{\citenamefont {Howell}(2003)}]{howell2003surface}%
  \BibitemOpen
  \bibfield  {author} {\bibinfo {author} {\bibfnamefont {P.~D.}\ \bibnamefont
  {Howell}},\ }\href@noop {} {\bibfield  {journal} {\bibinfo  {journal} {J.
  Eng. Math.}\ }\textbf {\bibinfo {volume} {45}},\ \bibinfo {pages} {283}
  (\bibinfo {year} {2003})}\BibitemShut {NoStop}%
\bibitem [{\citenamefont {Roberts}\ and\ \citenamefont
  {Li}(2006)}]{roberts2006accurate}%
  \BibitemOpen
  \bibfield  {author} {\bibinfo {author} {\bibfnamefont {A.~J.}\ \bibnamefont
  {Roberts}}\ and\ \bibinfo {author} {\bibfnamefont {Z.}~\bibnamefont {Li}},\
  }\href@noop {} {\bibfield  {journal} {\bibinfo  {journal} {J. Fluid Mech.}\
  }\textbf {\bibinfo {volume} {553}},\ \bibinfo {pages} {33} (\bibinfo {year}
  {2006})}\BibitemShut {NoStop}%
\bibitem [{\citenamefont {Thiffeault}\ and\ \citenamefont
  {Kamhawi}(2006)}]{thiffeault2006transport}%
  \BibitemOpen
  \bibfield  {author} {\bibinfo {author} {\bibfnamefont {J.-L.}\ \bibnamefont
  {Thiffeault}}\ and\ \bibinfo {author} {\bibfnamefont {K.}~\bibnamefont
  {Kamhawi}},\ }\href@noop {} {\bibfield  {journal} {\bibinfo  {journal} {arXiv
  preprint nlin/0607075}\ } (\bibinfo {year} {2006})}\BibitemShut {NoStop}%
\bibitem [{\citenamefont {Wray}\ \emph {et~al.}(2017)\citenamefont {Wray},
  \citenamefont {Papageorgiou},\ and\ \citenamefont {Matar}}]{wray2017reduced}%
  \BibitemOpen
  \bibfield  {author} {\bibinfo {author} {\bibfnamefont {A.~W.}\ \bibnamefont
  {Wray}}, \bibinfo {author} {\bibfnamefont {D.~T.}\ \bibnamefont
  {Papageorgiou}}, \ and\ \bibinfo {author} {\bibfnamefont {O.~K.}\
  \bibnamefont {Matar}},\ }\href@noop {} {\bibfield  {journal} {\bibinfo
  {journal} {SIAM J. Appl. Math.}\ }\textbf {\bibinfo {volume} {77}},\ \bibinfo
  {pages} {881} (\bibinfo {year} {2017})}\BibitemShut {NoStop}%
\bibitem [{\citenamefont {Shepherd}\ \emph {et~al.}(2024)\citenamefont
  {Shepherd}, \citenamefont {Boujo},\ and\ \citenamefont
  {Sellier}}]{shepherd2024general}%
  \BibitemOpen
  \bibfield  {author} {\bibinfo {author} {\bibfnamefont {R.~G.}\ \bibnamefont
  {Shepherd}}, \bibinfo {author} {\bibfnamefont {E.}~\bibnamefont {Boujo}}, \
  and\ \bibinfo {author} {\bibfnamefont {M.}~\bibnamefont {Sellier}},\
  }\href@noop {} {\bibfield  {journal} {\bibinfo  {journal} {J. Fluid Mech.}\
  }\textbf {\bibinfo {volume} {998}},\ \bibinfo {pages} {A32} (\bibinfo {year}
  {2024})}\BibitemShut {NoStop}%
\bibitem [{\citenamefont {Irgens}\ \emph {et~al.}(2019)\citenamefont {Irgens},
  \citenamefont {Irgens},\ and\ \citenamefont {Baumann}}]{irgens2019tensor}%
  \BibitemOpen
  \bibfield  {author} {\bibinfo {author} {\bibfnamefont {F.}~\bibnamefont
  {Irgens}}, \bibinfo {author} {\bibfnamefont {B.}~\bibnamefont {Irgens}}, \
  and\ \bibinfo {author} {\bibfnamefont {D.}~\bibnamefont {Baumann}},\
  }\href@noop {} {\emph {\bibinfo {title} {Tensor analysis}}}\ (\bibinfo
  {publisher} {Springer},\ \bibinfo {year} {2019})\BibitemShut {NoStop}%
\end{thebibliography}%

\end{document}


\title{Supporting Information: Thin-film drainage becomes singular at saddles }

\author{Simeon Djambov\,}
\affiliation{Laboratory of Fluid Mechanics and Instabilities, EPFL, Lausanne CH-1015, Switzerland}
\affiliation{LadHyX, CNRS, Ecole Polytechnique, Institut Polytechnique de Paris, Palaiseau, France}

\author{Alice Marcotte\,}
\affiliation{Institut Jean le Rond d’Alembert, CNRS, Sorbonne Université, Paris, France}

\author{Fran\c{c}ois Gallaire\,}
\affiliation{Laboratory of Fluid Mechanics and Instabilities, EPFL, Lausanne CH-1015, Switzerland}

\author{Pier Giuseppe Ledda\,}
\email{piergiuseppe.ledda@unica.it}
\affiliation{Department of Civil, Environmental Engineering and Architecture, University of Cagliari, 09123 Cagliari, Italy}

\date{\today}

\begin{abstract}

\end{abstract}

\maketitle

\setcounter{equation}{0}
\setcounter{figure}{0}
\setcounter{table}{0}
\setcounter{page}{1}
\makeatletter
\renewcommand{\theequation}{S\arabic{equation}}
\renewcommand{\thefigure}{S\arabic{figure}}

\section{Experimental details}

\textbf{Oil properties.}
Silicone oils, purchased from Carl Roth (CAS number: 63148-62-9), of kinematic viscosities 
$\nu = 350~\mathrm{mm}^2\,\mathrm{s}^{-1}$ and $\nu = 1000~\mathrm{mm}^2\,\mathrm{s}^{-1}$ (manufacturer specifications: 333–368 and 950–1050~$\mathrm{mm}^2\,\mathrm{s}^{-1}$, respectively) were employed in the experiments.
In both cases, the density at $25^\circ\mathrm{C}$ is {$\rho = 970 \pm 5 ~\mathrm{kg\,m}^{-3}$ 
and the surface tension is $\gamma = 21 \pm 0.2~\mathrm{mN\,m}^{-1}$ \cite{wang2026}}. 
\vspace{0.1cm}

\textbf{3D printing.}

\begin{figure}[!h]
    \centering
\includegraphics[width=0.5\linewidth]{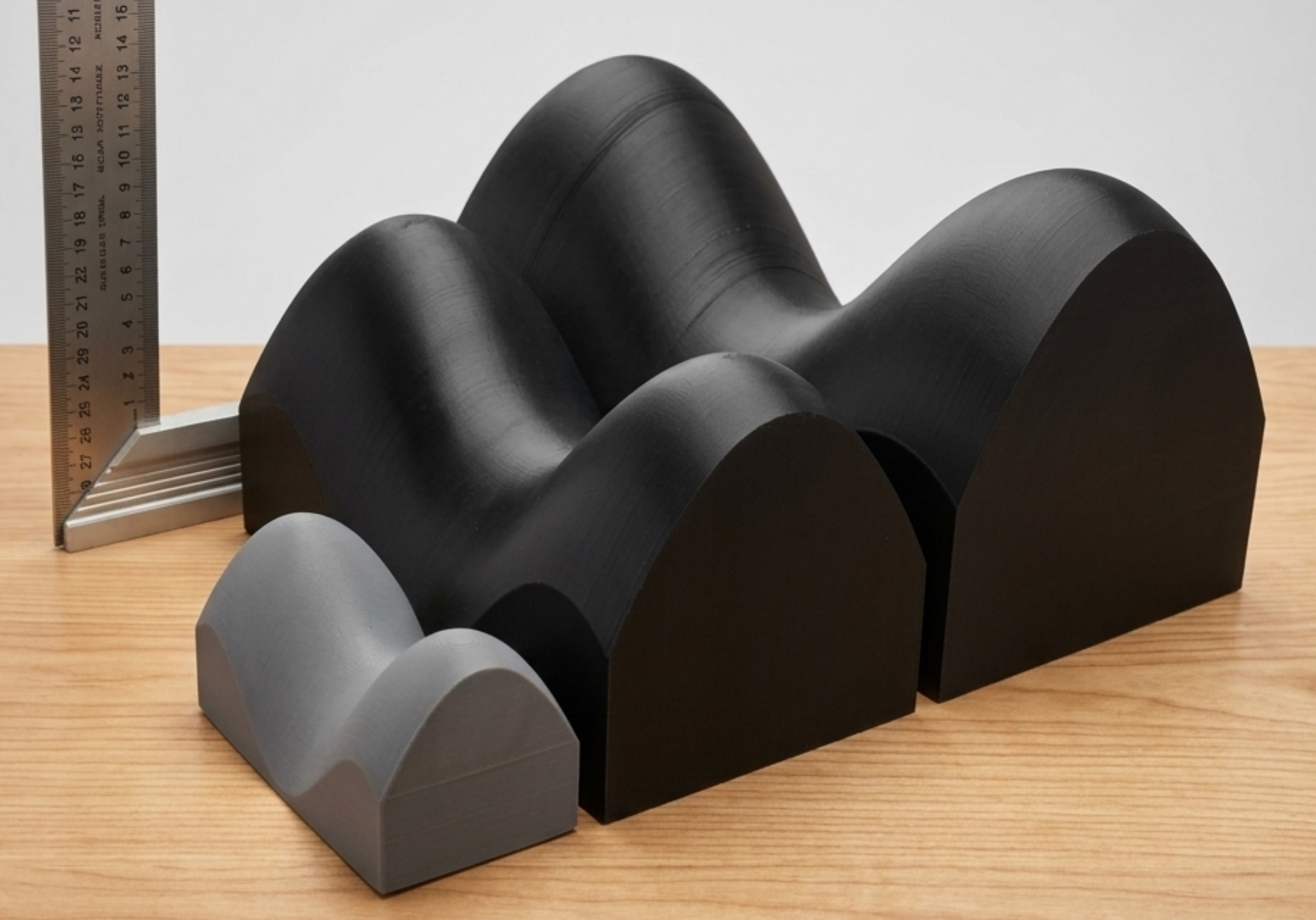}
    \caption{{Image of the three substrates used in the experiments. The grey substrate corresponds to $Bo = 102 \pm 1$, while the two black substrates correspond to $Bo = 408 \pm 5$ and $Bo = 725 \pm 9$.}}
    \label{fig:substrates}
\end{figure}

{
Three geometrically similar substrates were fabricated in-house with characteristic length scales $L = 15 \pm 0.025$, $30 \pm 0.1$, and $40 \pm 0.1~\mathrm{mm}$, yielding Bond numbers $Bo =  \rho g L^2 / \gamma=102 \pm 1$, $Bo = 408 \pm 5$, and $Bo = 725 \pm 9$, respectively (see Figure~\ref{fig:substrates}). The smallest substrate was produced by stereolithography (SLA), which provides a characteristic resolution of approximately $25~\mu\mathrm{m}$. Due to the limited build volume of the available SLA system, the two larger substrates were fabricated by fused filament fabrication (FFF), with a characteristic resolution of approximately $100~\mu\mathrm{m}$.
In all cases, the upper surface of the substrate is described by
$
\eta(x,y) = L - L\cos\!\left(\frac{y}{L}\right) - \frac{x^2}{2L}
$.}
\vspace{0.1cm}

\begin{figure}[!h]
    \centering
\includegraphics[width=0.8\linewidth]{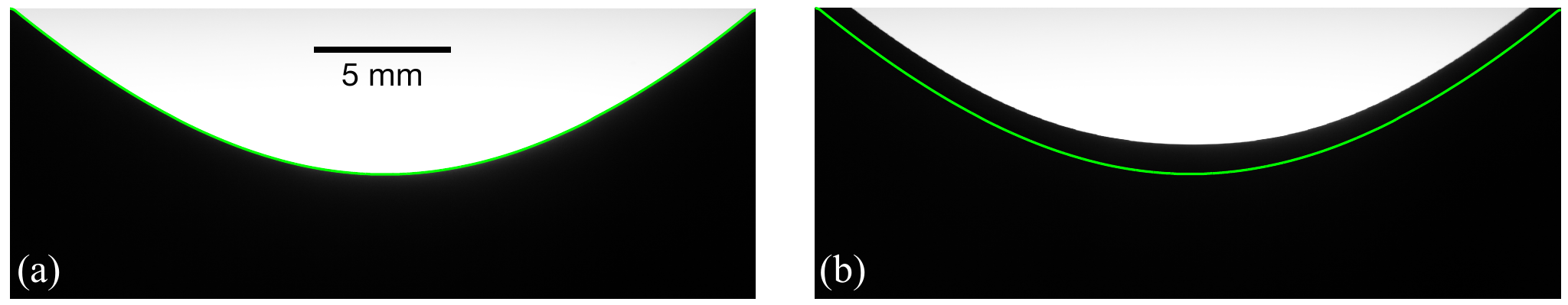}
    \caption{(a) Image of the dry substrate {($Bo =102$)}. (b) Image of the same substrate covered by a silicone oil film immediately after pouring. In both images, the field of view corresponds to the plane $x=0$, in the vicinity of the saddle point (small $y$). The green curve indicates the position of the solid--air interface of the dry substrate.}
    \label{fig:before_after_pouring}
\end{figure}

\textbf{Silicone oil deposition.}
A large quantity of silicone oil was poured onto the substrate, resulting in an initial condition corresponding to a relatively thick liquid layer, with a thickness of a few mm at the saddle region, from which drainage subsequently develops (see Figure \ref{fig:before_after_pouring}). 
\vspace{0.1cm}

\begin{figure}[!h]
    \centering
\includegraphics[width=0.6\linewidth]{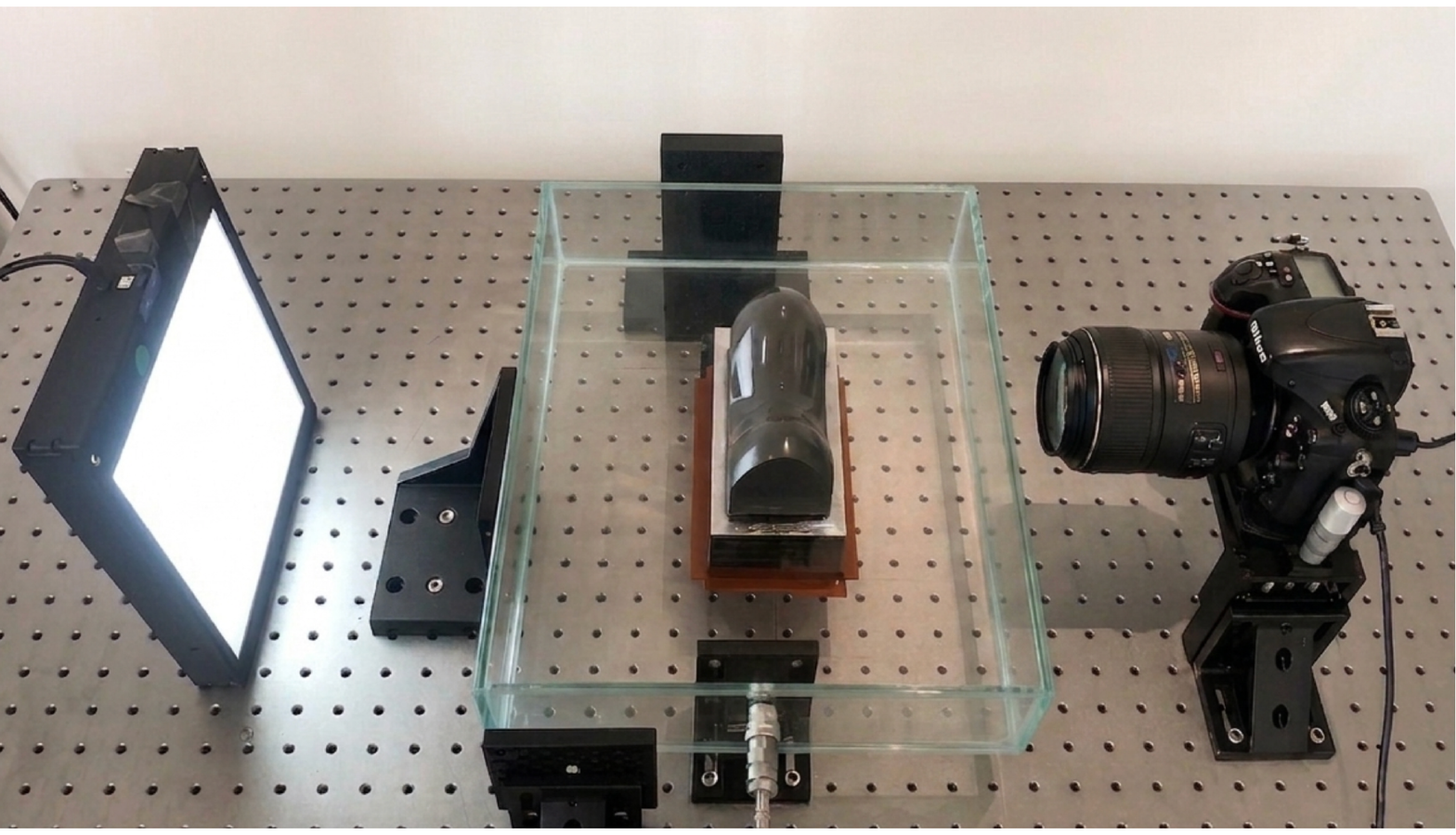}
    \caption{Image of the experimental set-up.}
    \label{fig:set-up}
\end{figure}

\textbf{Imaging method.} The draining film was imaged using a Nikon D800 camera equipped with an AF-S Micro Nikkor 105~mm f/2.8G IF-ED macro-lens (spatial resolution $\sim$ 5 $\mu$m/pixel). A uniform LED panel provided back-illumination (see Figure \ref{fig:set-up}). The camera was focused on the symmetry plane $x=0$ near the saddle, and the optical axis of the camera was oriented horizontally and intersected the location of the minimum of the valley. A reference image of the dry substrate was first acquired to determine the position and geometry of the substrate in the imaging plane. Time-lapse imaging was initiated at the moment the silicone oil was poured onto the substrate, with the first frame defining the initial time $t=0$. 

Movies were recorded by acquiring one image every 5 to 10~s, depending on the oil viscosity and the substrate size.
\vspace{0.1cm}

\begin{figure}[t]
    \centering
\includegraphics[width=0.6\linewidth]{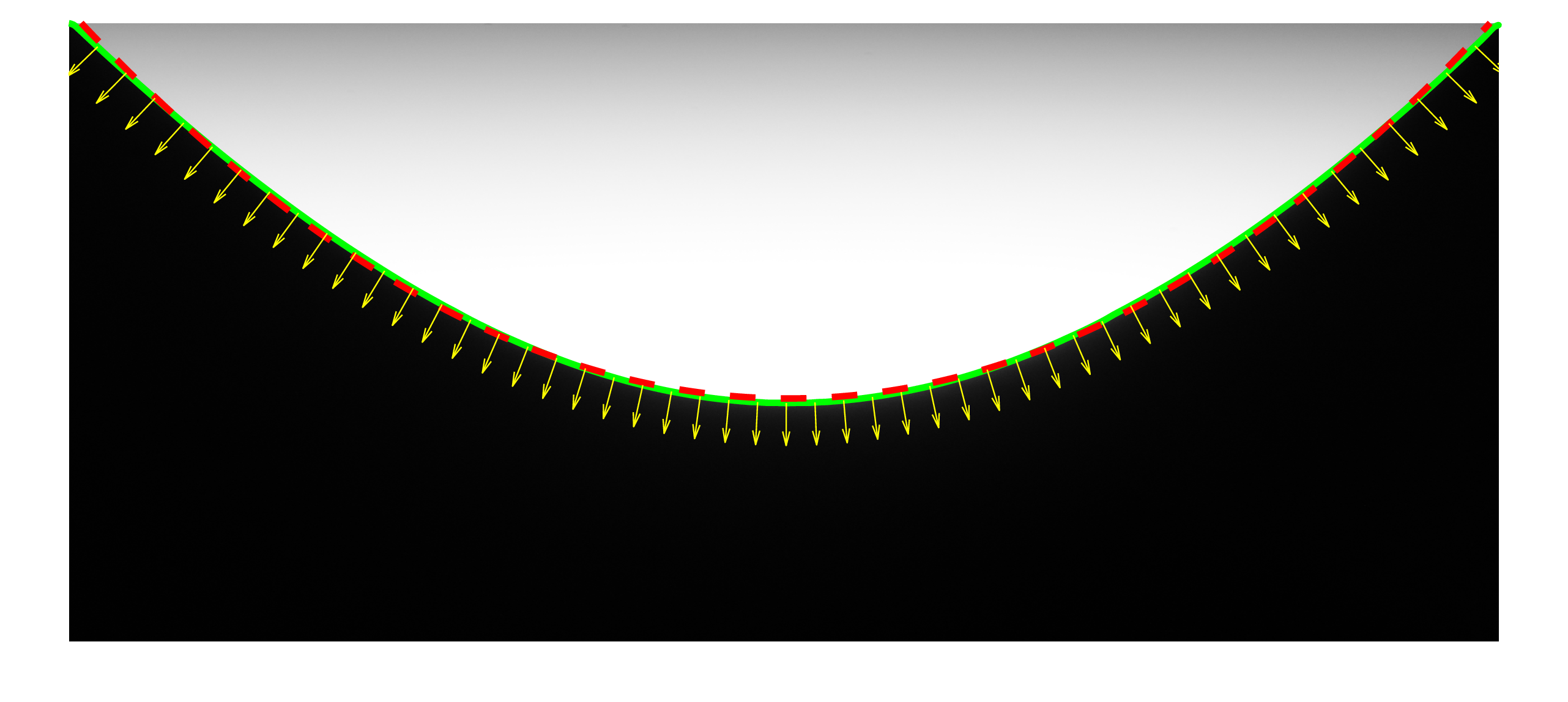}
    \caption{Image of the dry substrate {($Bo =102$)}. The green curve indicates the location of the solid–air interface, obtained by image thresholding. The red dotted line corresponds to the best parabolic fit. The yellow arrows indicate the local surface normal vectors in the $x=0$ plane.}
    \label{fig:fit_parabola}
\end{figure}

\textbf{Thickness extraction.} The image of the dry substrate is first converted to grayscale.
The solid-air interface is then detected by binarization using a fixed intensity threshold. Specifically, at each horizontal position, the interface is taken as the uppermost detected dark pixel, yielding a discrete curve $y_{\mathrm{sub}}(x)$ which is subsequently interpolated and smoothed.
To account for possible misalignment in the measurement plane between the substrate and the imaging axes, the extracted interface is fitted in a rotated coordinate system. 
Specifically, a rotation angle $\theta$ is optimized by minimizing the mean squared deviation between the data and a parabolic fit of the form
$
y' = a x'^2 + b x' + c.
$
The optimal parabola is obtained via a least-squares fit in the rotated frame, and the resulting curve is mapped back to the original image coordinates (see Fig.~\ref{fig:fit_parabola}).
The quadratic coefficient provides an independent estimate of the geometric length scale through $a = 1/(2L)$. 
Uncertainties in $L$ are evaluated by varying the intensity threshold used for interface detection, leading to variations of up to 5\% in the value of $L$ inferred from the procedure.
Local normal vectors to the substrate are then computed from the tangent of the fitted parabola in the rotated frame and rotated back to the image frame.
The remaining out-of-plane perspective distortions, associated with rotations about the other two axes, were neglected. This is justified by the close proximity of the camera to the substrate and by the small field of view.

\begin{figure}[!h]
    \centering
\includegraphics[width=0.7\linewidth]{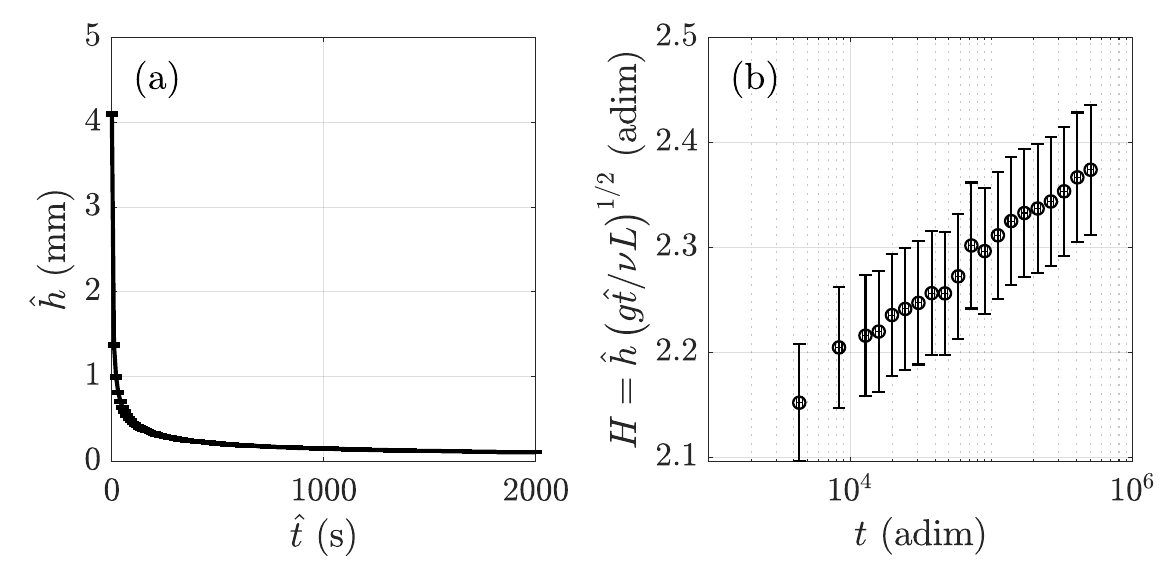}
    \caption{(a) Liquid film thickness $\hat h$ as a function of time $\hat t$ measured in the plane $x=0$ along the local normal to the substrate at the position $y=0$, for $Bo=725$ and $\nu=1000~\mathrm{mm}^2\mathrm{s}^{-1}$. (b) The corresponding dimensionless quantity $H = \hat h (g\hat t /\nu L)^{1/2}$ as a function of time.}
    \label{fig:h_vs_t}
\end{figure}

In the following, $(x_S,y_S)$ denote coordinates in the image plane. 
The image horizontal axis corresponds to the physical $y$-direction, while the vertical axis corresponds to the physical $z$-direction, within the $x=0$ plane. 
Using the local normal vectors, the time-lapse movies are resliced along directions normal to the substrate. 
At each position $(x_S,y_S)$ sampled uniformly in the vicinity of the parabola minimum, a line is defined along the corresponding unit normal vector $\mathbf{n}(x_S)$. 
Image intensities are then obtained by bilinear interpolation along trajectories of the form
$
\mathbf{r}(s) = (x_S, y_S) + s\,\mathbf{n}(x_S),
$
for each frame of the time-lapse sequence. 
The reslicing procedure transforms the time-lapse movies into spatio-temporal intensity maps, where the horizontal axis corresponds to time and the vertical axis represents image intensity sampled along the local normal direction to the substrate. For each time and position along the substrate, the liquid--air interface is detected as the location of the maximum gradient of intensity along the vertical direction of the resliced image.
The solid--liquid interface corresponds to $s=0$, such that the local film thickness is directly obtained as the distance between the detected interface and the substrate along the normal direction. With this procedure, we experimentally reconstruct the film thickness $\hat h(\hat x=0,\hat y,\hat t)$, defined along the local normal direction to the substrate in the $x=0$ plane, see Figure \ref{fig:h_vs_t}a. Note that measurements are discarded once the thickness $\hat h$ becomes comparable to the substrate resolution.

The uncertainty associated with the measured film thickness $\hat h$ arises primarily from the calibration factor used to convert pixel distances into physical lengths, as discussed above. An additional contribution stems from the determination of the liquid--air interface position from the maximum of the intensity gradient in the resliced images. Owing to the use of interpolation around the gradient maximum, this localization uncertainty is subpixel. In practice, this contribution remains small compared with the calibration uncertainty for all retained measurements. Accordingly, the total uncertainty on $\hat h$ is taken to be dominated by the scale uncertainty.

\vspace{0.1cm}

\textbf{Comparison with theory.}
For comparison with the theoretical prediction, we consider the dimensionless quantity
\[
H = \hat h \left(\frac{g\,\hat t}{\nu L}\right)^{1/2},
\]
where $\hat h$ is the measured physical thickness and $\hat t$ is the physical time. 

Since the theoretical prediction concerns the dependence of $H$ on $\log t$, while the time-lapse movies were acquired at constant time intervals, the experimental data are resampled onto a logarithmic time grid. To this end, the variable $\log t$ is divided into $N_{\mathrm{bins}}$ equally spaced bins (with $N_{\mathrm{bins}}\sim \sqrt{N}$, where $N$ is the number of data points). Within each bin, the mean value of $H$ is computed, yielding a binned quantity $H_{\mathrm{binned}}$ associated with the bin-center time.

{The uncertainty on $H$ arises from three independent contributions. The first is inherited from the uncertainty on the thickness measurement $\hat h$, discussed above. Since $H$ depends linearly on $\hat h$, this contribution gives a relative uncertainty identical to that of $\hat h$, $\frac{\delta H_{\hat h}}{H}=\frac{\delta \hat h}{\hat h}.$}
{The second contribution originates from the uncertainty on the kinematic viscosity $\nu$. Since $H\propto \nu^{-1/2}$, standard error propagation gives
$\frac{\delta H_{\nu}}{H}=\frac{1}{2}\frac{\delta\nu}{\nu}.$}
{The third contribution originates from the logarithmic time binning and reflects the statistical dispersion of the data within each bin. It is estimated from the standard error of the mean,
$\delta H_{\mathrm{bin}}=\frac{\sigma_H}{\sqrt{N_{\mathrm{bin}}}}$,
where $\sigma_H$ is the standard deviation of the values of $H$ in the bin and $N_{\mathrm{bin}}$ is the number of samples in that bin.}
{Assuming these three sources of uncertainty to be independent, the total uncertainty on the binned quantity is obtained from the quadratic sum
$\delta H_{\mathrm{tot}}=\sqrt{\delta H_{\hat h}^{\,2}+\delta  H_{\nu}^{\,2}+\delta H_{\mathrm{bin}}^{\,2}}.$}
{The dimensionless time is defined as $t=\frac{gL\,\hat t}{\nu}$, and is therefore also affected by the uncertainty on the viscosity. Since $t\propto \nu^{-1}$, its relative uncertainty is $\frac{\delta t}{t}=\frac{\delta\nu}{\nu}.$}
{The resulting data are represented as $H_{\mathrm{binned}}$ versus the logarithmically resampled dimensionless time, with vertical error bars corresponding to $\delta H_{\mathrm{tot}}$ and horizontal error bars corresponding to $\delta t$ (see Fig.~\ref{fig:h_vs_t}b).}

\begin{figure}
    \centering
    \begin{subfigure}[t]{0.34\textwidth}
    \caption{\raggedright}
    \includegraphics[width=\linewidth]{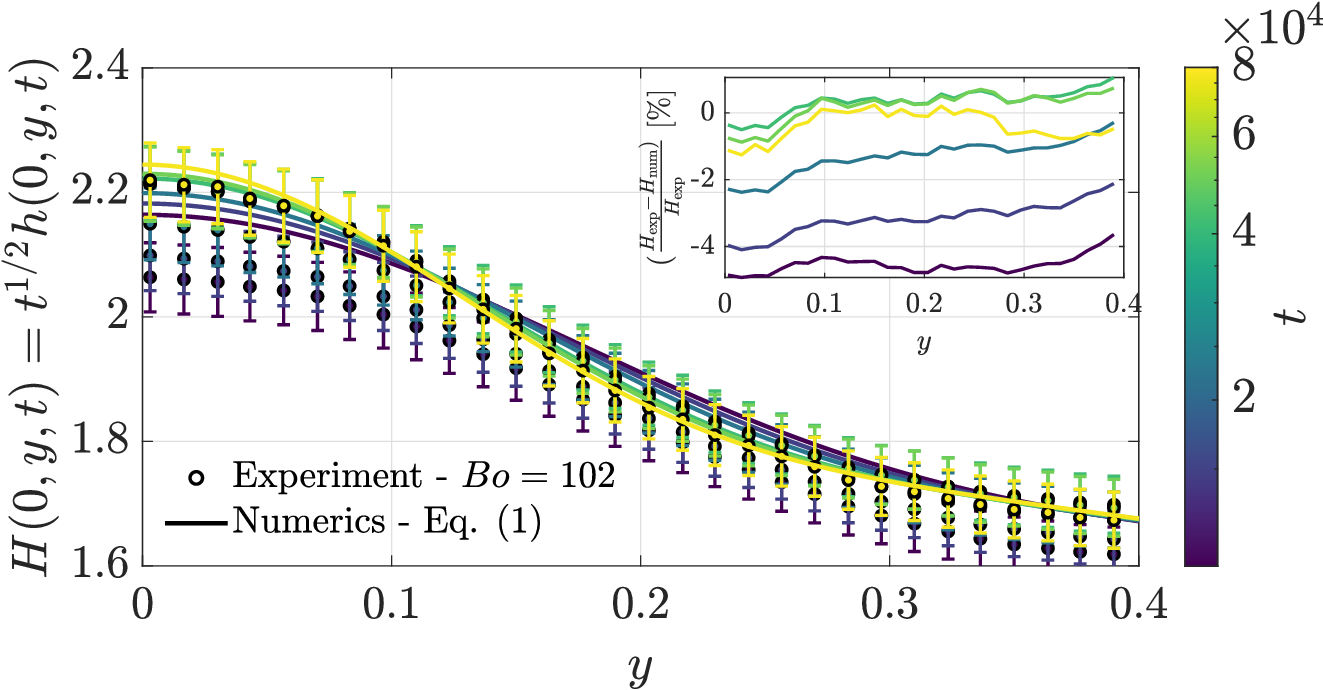}
    \end{subfigure}
    \begin{subfigure}[t]{0.34\textwidth}
    \caption{\raggedright}
    \includegraphics[width=\linewidth]{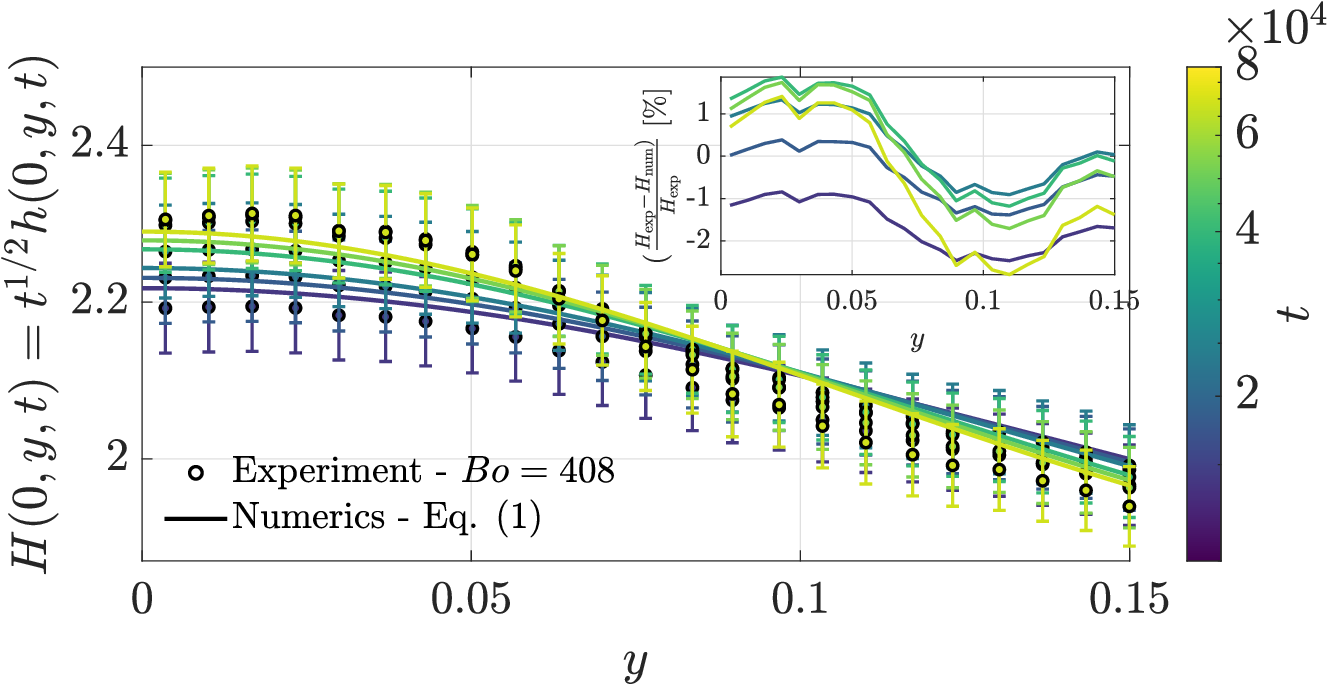}
    \end{subfigure}
    \begin{subfigure}[t]{0.28\textwidth}
    \caption{\raggedright}
    \includegraphics[width=\linewidth]{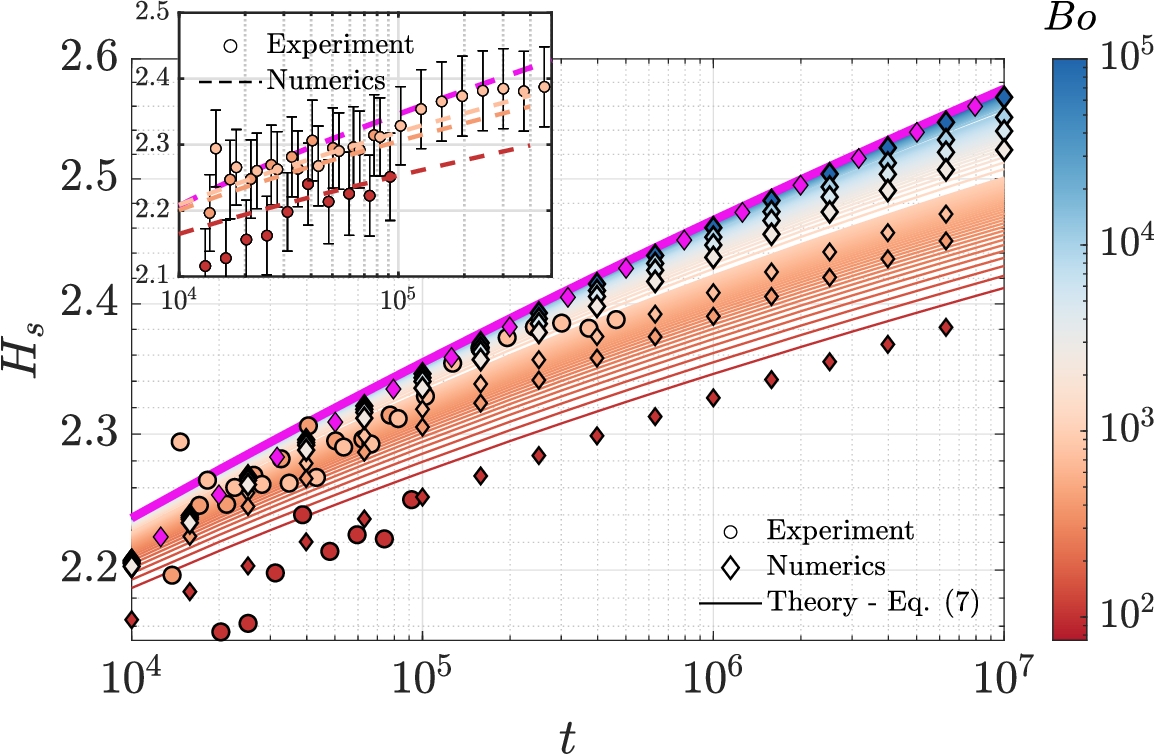}
    \end{subfigure}
    \caption{{
Additional experiments with the lower-viscosity silicone oil,
\(\nu=350\pm17\,\mathrm{mm^2\,s^{-1}}\).
(a,b) Rescaled thickness profiles \(H=t^{1/2}h\) measured along \(x=0\) for
(a) \(Bo=102\) and (b) \(Bo=408\).
Symbols denote experimental measurements and solid lines the corresponding
numerical solutions; colors indicate increasing nondimensional time.
(c) Rescaled saddle thickness \(H_s=H(0,0)\) versus nondimensional time.
Diamonds denote numerical solutions with \(h_0=0.1\) for, from bottom to top,
\(Bo=102,408,725\), corresponding to the experimental cases, followed by
\(Bo=2500,5000,10^4,\) and \(10^5\).
Circles denote experimental measurements with
\(\nu=0.35\times10^{-3}\,\mathrm{m^2\,s^{-1}}\) for
\(Bo=102,408,\) and \(725\).
Colored solid lines are the finite-capillarity saddle-layer predictions,
while the magenta solid line is the capillary-free limit.
The inset shows the three experimental datasets with error bars together
with the corresponding numerical solutions (dashed lines), while the magenta
dashed line denotes the numerical solution without capillarity.}}
    \label{fig:350cst}
\end{figure}

{
\section{Additional experimental data}

To assess the robustness of the observed drainage mechanism, we report experiments using silicone oil with different kinematic viscosity
\(\nu=350\pm17\,\mathrm{mm^2\,s^{-1}}\). Figure~\ref{fig:350cst}(a,b) shows the rescaled thickness profiles
\(H=t^{1/2}h\) measured along \(x=0\), together with the corresponding numerical solutions for $Bo=102$ and $Bo =  408$. As for the \(\nu=1000\,\mathrm{mm^2\,s^{-1}}\) experiments, the profiles develop a localized maximum near the saddle while approaching the outer drainage behavior farther away, for large time.
Figure~\ref{fig:350cst}(c) reports the rescaled saddle thickness
\(H_s=H(0,0)\) as in Fig.~3 of the Letter. The same time and Bond-number dependence is observed.
}

\section{Lubrication equation in generalized coordinates}

In this Section, we recall differential geometry concepts \cite{irgens2019tensor} to define the lubrication equation on a generic curved substrate, whose derivation is well-established in the literature \cite{roy2002lubrication,thiffeault2006transport,wray2017reduced,shepherd2024general}. 

\subsection{Differential geometry concepts}
We introduce a Cartesian reference frame $({x},{y},{z})$. A fluid film of thickness ${h}$ lies on a substrate of height $\eta$. 
The substrate is identified via a position vector $\boldsymbol{X}\left(x^{\{1\}}, x^{\{2\}}\right)$, where ($x^{\{1\}},x^{\{2\}}$) are the local coordinates used to parameterize the surface.
We introduce a local reference frame with two directions lying on the substrate ${\mathbf{e}}_{i}=\partial_i \boldsymbol{X}$ ($i=1,2$), not necessarily orthogonal, and the normal-to-the-substrate coordinate vector ${\mathbf{e}}_{3}=\frac{\mathbf{e}_{1}\times\mathbf{e}_{2}}{|\mathbf{e}_{1}\times\mathbf{e}_{2}|}$.  
We introduce the {$2\times2$ symmetric} metric tensor {$\mathbb{G}_{i j}$ and the square root of the determinant of the metric tensor $w$, related to the area element on the surface $\mathrm{d}A$ through $dA=w \mathrm{d}x^{\{1\}}\mathrm{d}x^{\{2\}}$ :
\begin{equation}
\mathbb{G}_{i j}=\mathbf{e}_{i} \cdot \mathbf{e}_{j}=\mathbb{G}_{j i}, \quad w=\left(\operatorname{det} \mathbb{G}_{i j}\right)^{1 / 2}.
\end{equation}
We define the second fundamental form and the curvature tensor, whose components respectively read, {following Einstein's notation for the summation}:
\begin{equation}
\mathbb{S}_{i j}=\partial_i \mathbf{e}_{j} \cdot {\mathbf{e}}_3, \quad
\mathbb{K}_{i}^{\{j\}}=\mathbb{S}_{i k}\mathbb{G}^{\{kj\}},
\end{equation}
where $\mathbb{G}^{\{ij\}}$ is the inverse metric tensor, i.e. $\mathbb{G}^{\{ij\}}=\mathbb{G}_{ij}^{-1}$.
The mean $\mathcal{K}$ and the Gaussian $\mathcal{G}$ curvatures read
$\mathcal{K}=\operatorname{tr}\mathbb{K}$ and $\mathcal{G}=\operatorname{det} \mathbb{K}$, 
respectively.
A generic vector $\mathbf{f}$ can be written in terms of its covariant and contravariant base, i.e. $\mathbf{f}=f^{\{i\}}\mathbf{e}_i=f_i\mathbf{e}^{\{i\}}$, where $\mathbf{e}^{\{i\}}$ is the covector defined as 
${\mathbf{e}}^{\{i\}} \cdot {\mathbf{e}}_{j}=\delta_{ij}$.
The two contravariant components, parallel to the substrate, of the gravity vector read $g_t^{\{i\}}=\mathbf{g}\cdot \mathbf{e}^{\{i\}}$, while the normal one reads $g_3=\mathbf{g}\cdot {\mathbf{e}}_3$.
The gradient of a scalar function $f$ and the divergence of a generic vector $\mathbf{f}=f^{\{i\}}\textbf{e}_i$ respectively read \citep{irgens2019tensor}:
\begin{equation}
\nabla f=\mathbb{G}^{\{i j\}}{\partial_j f} \, \mathbf{e}_{i}=\partial^{\{i\}} f\, \mathbf{e}_{i}, \quad
\nabla \cdot\mathbf{f} =w^{-1} \partial_{i}(w f^{\{i\}}).
\end{equation}

\subsection{Complete lubrication model in generalized coordinates}

{We non-dimensionalize with the characteristic length $L$ of the substrate and with the characteristic time $\tau={\nu}/({ g L})$.
Upon non-dimensionalization, the equation in coordinate-free form reads \citep{roy2002lubrication,howell2003surface,roberts2006accurate,thiffeault2006transport}:
\begin{equation}
\begin{aligned}
(1-\mathcal{K}h+\mathcal{G}h^{2})\frac{\partial h}{\partial t}&+\frac{1}{3Bo} \nabla \cdot\left[h^{3} \left( \nabla \tilde{\kappa}-\frac{1}{2} h(2\mathcal{K} \mathbb{I}-\mathbb{K}) \cdot \nabla \mathcal{K}\right)\right] \\
&+\frac{1}{3} \nabla \cdot\left[h^{3} \left({ \boldsymbol{g}}_{t}-  h\left(\mathcal{K} \mathbb{I}+\frac{1}{2} \mathbb{K}\right) \cdot {\boldsymbol{g}}_{t}+  {\mathrm{g}}_{3} \nabla h\right)\right]=0,
\label{eq:lubr_compl1}
\end{aligned}
\end{equation}
where $\tilde{\kappa}=\mathcal{K}+(\mathcal{K}^{2}-2\mathcal{G})h+ \nabla^2h$ is the free-surface curvature, $Bo=(\rho g L^2)/\gamma$ is the Bond number and ${\boldsymbol{g}}_{t}$ and ${\mathrm{g}}_{3}$ are the gravity vector components tangent and normal to the substrate, respectively, normalized by the acceleration of gravity.}
Equation \eqref{eq:lubr_compl1} is the full model from which Eq. (1) in the main text is obtained by grouping the leading drainage, hydrostatic, and capillary fluxes.
{The terms in the first and second bracket represent the flux induced by capillary and gravity effects, respectively. At leading order, capillary flow is induced by variations of the mean curvature of the substrate $\mathcal{K}$ and leads to film thinning and thickening in the neighborhood of local maximum and minimum values of the curvature \citep{roy2002lubrication}. In addition, we have (i) free-surface curvature variations and (ii) higher order terms of the substrate curvature. Gravity-induced fluxes are instead related to the gravity components tangential to the substrate $\boldsymbol{g}_t$ and hydrostatic pressure gradients along the thin film.
The purely gravity-driven problem is obtained by assuming $Bo \gg 1$ so that terms related to surface tension disappear from the problem.}

\subsection{Monge parameterization of a generic substrate}
\label{appamonge}

In the \emph{Monge parameterization}, the substrate position vector is defined as
$\boldsymbol{X}=(x,y, \eta (x,y))$, with $x^{\{1\}}=x, x^{\{2\}}=y$.
For ease of notation, we indicate derivatives of substrate-related quantities with subscripts. The tangent and normal vectors to the substrate read:
\begin{equation}
\mathbf{e}_{1}=\partial_{1} \boldsymbol{X}=\left[
1,  0,  \eta_x\right], \quad \mathbf{e}_{2}=\partial_{2} \boldsymbol{X}=\left[
0 , 1 , \eta_y\right], \quad {\mathbf{e}}_{3}=\frac{\mathbf{e}_{1}\times\mathbf{e}_{2}}{|\mathbf{e}_{1}\times\mathbf{e}_{2}|}=\frac{1}{n}[
-\eta_x,  -\eta_y, 1],
\end{equation}
with $n=\left(1+\eta_x^2+\eta_y^2\right)^{1/2}$.
The metric and inverse metric tensors read, respectively:
\begin{equation}
\begin{aligned}
\mathbb{G}_{i j}=\mathbf{e}_{i} \cdot \mathbf{e}_{j}=\left(\begin{array}{cc}
1+\left(\eta_x\right)^{2} & \eta_x \eta_y \\
\eta_x \eta_y & 1+\left(\eta_y\right)^{2}
\end{array}\right), \quad \mathbb{G}^{\{i j\}}=\mathbf{e}^{\{i\}} \cdot \mathbf{e}^{\{j\}}=\frac{1}{n^{2}}\left(\begin{array}{cc}
1+\left(\eta_y\right)^{2} & -\eta_x \eta_y \\
-\eta_x \eta_y & 1+\left(\eta_x\right)^{2}
\end{array}\right).
\end{aligned}
\end{equation}

The curvature tensor reads:
\begin{equation}
\mathbb{K}_{i}^{\{j\}}=\frac{1}{n^{3}}\left(\begin{array}{c c}
\left(1+\left(\eta_y\right)^{2}\right) \eta_{xx}-\eta_x \eta_y \eta_{xy}&\left(1+\left(\eta_x\right)^{2}\right) \eta_{xy}-\eta_x \eta_y \eta_{xx} \\
\left(1+\left(\eta_y\right)^{2}\right) \eta_{xy}-\eta_x \eta_y \eta_{yy}&\left(1+\left(\eta_x\right)^{2}\right) \eta_{yy}-\eta_x \eta_y \eta_{xy}
\end{array}\right),
\end{equation}
the mean and Gaussian curvature thus read:
\begin{equation}
\mathcal{K}=\frac{1}{n^{3}}\left(\left(1+\left(\eta_y\right)^{2}\right) \eta_{xx}-2 \eta_x \eta_y \eta_{xy}+\left(1+\left(\eta_x\right)^{2}\right) \eta_{yy}\right), \quad
\mathcal{G}=\frac{1}{n^{4}}\left(\eta_{xx} \eta_{yy}-\left(\eta_{xy}\right)^{2}\right).
\end{equation}

The gravity vector, which reads ${\boldsymbol{g}}=[0,0,-1]$ in the $(x,y,z)$ coordinate system, is projected on the contravariant base:
\begin{equation}
{g}_{{t}}^{\{1\}}={\boldsymbol{g}} \cdot \mathbf{e}^{\{1\}}=-\eta_x/n^{2},\quad
{g}_{{t}}^{\{2\}}={\boldsymbol{g}} \cdot \mathbf{e}^{\{2\}}=-\eta_y /n^{2}, \quad{g}_{3}={\boldsymbol{g}} \cdot {\mathbf{e}}_{3}=-1 / n.
\end{equation}

These equations close the lubrication model.
We considered two different substrates, i.e. (i) the periodic saddle case $\eta=1-\cos y - x^2/2$, and (ii) the hyperbolic paraboloid $\eta=y^2/2 - x^2/2$.

\section{Numerical details}

\subsection{Full model}

Numerical simulations of the full model are performed in COMSOL Multiphysics via the \textit{General Form PDE} interface, where the equation is solved in its conservation form. The implementation is based on the definition of the substrate height $\eta$ in its analytical form, from which all derivatives and differential-geometry quantities are automatically evaluated after discretization. Quadratic Lagrangian elements are exploited for the numerical discretization while the time-marching is performed via the built-in BDF solver, with a relative tolerance of $10^{-5}$. Leveraging symmetry, the model is solved for ($x>0,y>0$), imposing zero flux conditions at $x=0$ and $y=0, \pi$. The outlet condition is imposed via a sponge that relaxes the thickness to zero. In the periodic saddle case, the sponge extends from $x=10$ to the outlet boundary at $x=12$. A convergence study is performed: the model is considered at convergence if further mesh refinement corresponds to a relative variation of the saddle thickness smaller than $0.1\%$, for fixed and large times. Note that the bottleneck is related to the mesh size at $y=0$, which requires, for large time, mesh sizes close to the saddle of at most $0.0005$, in dimensionless form, to ensure the required convergence. 

\subsection{Drainage problem}

Drainage profiles were computed within a Python routine using the method of characteristics on analytically defined surfaces. The equations are solved for $f := H_d^2$. Characteristic curves are initialized from a regular grid of starting points $(\Delta<x_0<5, 0<y_0<5)$, with $\Delta=0.00025$, and integrated forward in the direction of steepest descent using the library \emph{scipy.integrate.solve$\_$ivp} with a Dormand–Prince Runge–Kutta method of order 5(4), with relative and absolute tolerances of $10^{-7}$ and $10^{-10}$, respectively. Integrating backward, each trajectory terminates at the maximum height, here denoted as the pole, where the surface gradient magnitude drops below a fixed threshold ( $10^{-9}$). Along each characteristic, we solve a coupled system of ordinary differential equations for a particular solution $f_p(s)$ and its homogeneous counterpart $f_h(s)$, both initialized with unity.
For the saddle $\eta=1-\cos y -x^2/2$, the solution on the maxima is known \cite{ledda2022gravity} and used to rescale the thickness, which eliminates dependence on  arbitrary initial values. 

\subsection{Theoretical saddle solution}

The numerical matching procedure of the saddle problem with the drainage solution is performed in MATLAB. We employ the boundary-value problem solver \textit{bvp5c}, with an absolute tolerance of $10^{-8}$, with 20000 points, chosen after a convergence analysis. As boundary conditions, we impose zero derivative at $\zeta=0$. As discussed in the main text, the natural matching with the logarithmic
drainage profile for $\zeta \gg 1$ allows for a numerical matching at one point $\zeta_m$. The logarithmic solution becomes singular at $\zeta_b = \exp(C)t^{1/4}$. As discussed in the main text, we impose the second boundary condition at $\zeta_{m}$, $\mathcal H(\zeta_{m})=\sqrt{C+ \frac{1}{4}\log t-\log(\zeta_{m})}$. A variation of the matching point between $0.5 \zeta_b<\zeta_m<0.9 \zeta_b$ leads to a saddle thickness value constant up to four significant digits.
For the finite Bond case, the procedure is analogous with, in addition, zero third derivative at $\zeta=0$ and the matching is performed both for $\mathcal H(\zeta_{m})$ and $\mathcal H'(\zeta_{m})$, for the same $\zeta_m$.

\section{Drainage solution: method of characteristics}

We recast the drainage problem through the Monge parameterization:
\begin{equation}
h_t-\frac{1}{3 ({1+\eta_x^2+\eta_y^2})^{1/2}} {\nabla}_{xy} \cdot\left[\frac{h^3 {\nabla}_{xy} \eta}{({1+\eta_x^2+\eta_y^2})^{1/2}}\right]=0,
\end{equation}
Upon considering the large-time evolution, at leading order,
\begin{equation}
H_d+\frac{2}{3({1+\eta_x^2+\eta_y^2})^{1/2}}{\nabla}_{xy}\cdot\left[\frac{H_d^3{\nabla}_{xy} \eta}{({1+\eta_x^2+\eta_y^2})^{1/2}}\right]=0.
\end{equation}

By expanding the divergence, we obtain
\begin{equation}
 {\nabla}_{xy} \eta \cdot {\nabla}_{xy} H_d^2+w\left(\frac{2}{3}\mathcal{K} H_d^2+w\right)=0,
\label{eq:outer-drainage-pde}
\end{equation}
with the mean curvature $\mathcal{K}$. This equation suggests that the characteristic lines follow the gradient lines of the topography, as the advection of $H_d^2$ is precisely ${\nabla}_{xy} \eta$.
We solve the problem with the method of characteristics \cite{bers1964partial}. We introduce the curvilinear coordinate, descending along a gradient line, $s$ such that
\begin{equation}
    \begin{aligned}
\frac{\mathrm{d} x}{\mathrm{~d} s} & =-\frac{\eta_x}{({\eta_x^2+\eta_y^2})^{1/2}}, \quad \frac{\mathrm{~d} y}{\mathrm{~d} s}=-\frac{\eta_y}{({\eta_x^2+\eta_y^2})^{1/2}}, \\
\frac{\mathrm{~d} H_d^2}{\mathrm{~d} s} & =\frac{ w}{({\eta_x^2+\eta_y^2})^{1/2}}\left((2 / 3)\mathcal{K} H_d^2+w\right).
\end{aligned}
\end{equation}

Characteristic (gradient) lines emanate from the stationary points (maxima) of the topography. ``Draining'' maxima are those of strictly negative mean curvature $\mathcal{K}=\mathcal{K}_p<0$, where $\mathcal{K}_p$ is the curvature at the maximum of the substrate (in the periodic saddle case, $\mathcal K_p=\mathcal K(0,y_0=\pm (2n+1) \pi),\,n=0,1,2, \cdots $), where the thickness has been determined as $H_d^2= -3/(2\mathcal{K}_p)$ \cite{ledda2022gravity}. 

\section{Drainage solution approaching the principal descending direction}
Approaching $y=0$, the solution $H_d$ displays a singular behavior of the form
\begin{equation}
H_d(x,y\sim0)\sim\sqrt{C(x)-B(x)\ln|y|}.\label{eq2bis:outer-in}
\end{equation}
This can be best explained on the symmetry line $x=0$, where the drainage problem reads
\begin{equation}
\eta_y\dfrac{\mathrm{d}H_d^2}{\mathrm{d}y}+w\left(\dfrac{2}{3}\mathcal{K}H_d^2+w\right)=0\label{eq2bis:outer-x0}.
\end{equation}
As $y$ goes to $0$, approaching the saddle point of zero mean curvature, $\mathcal{K}=0$, the second term vanishes, while the other two terms with $\eta_y\sim y$ and $w\to1$, lead to the singular close-to-the-saddle behavior, as explained in the main text. 

Instead, by expanding eq.~\eqref{eq:outer-drainage-pde} for small $y$,
\begin{equation}
\eta_x\dfrac{\partial H_d^2}{\partial x}+\eta_{yy}|_0 y\dfrac{\partial H_d^2}{\partial y}+\dfrac{2}{3}\dfrac{\eta_{xx}+(1+\eta_x^2)\eta_{yy}|_0}{1+\eta_x^2}H_d^2+1+\eta_x^2+\mathcal{O}[y^2]=0,\label{eq2bis:outer-smally}
\end{equation}
and injecting eq.~\eqref{eq2bis:outer-in} as an ansatz, two ordinary differential equations for $B$ and $C$ as functions of $x$ may be identified,
\begin{subequations}\label{eq2bis:sing-drain}
\begin{align}
\eta_x B'+\dfrac{2}{3}\dfrac{\eta_{xx}+(1+\eta_x^2)\eta_{yy}|_0}{1+\eta_x^2}B&=0,\\
\eta_x C'-\eta_{yy}|_0 B+\dfrac{2}{3}\dfrac{\eta_{xx}+(1+\eta_x^2)\eta_{yy}|_0}{1+\eta_x^2}C+1+\eta_x^2&=0,
\end{align}
\end{subequations}
with initial conditions for $B(0)$ and $C(0)$. The first equation cancels all terms multiplying $\ln|y|$, while the second one ensures that all other terms, independent of $y$, vanish. While $B(0)$ is determined directly by the geometry close to the saddle, $C(0)$ is a constant of integration along the inflow characteristic, keeping information of the far field, as explained in the next section. Note that $\eta_x$, $\eta_{xx}$, and $\eta_{yy}|_0$ are all evaluated at $y=0$, but remain functions of $x$.

\section{Method for determining the constants in the close-to-the-saddle behavior of the drainage solution}

The purpose of this section is to compute the constant $C$ entering the close-to-the-saddle logarithmic profile of the main text.
Let us consider the characteristic line along $0\leq y\leq y_0=\pi$, with $x=0$, that reaches the saddle point at the origin. Along this characteristic, the drainage problem reads
\begin{equation}
    \frac{\mathrm{d} H_d^2}{\mathrm{d} y}
    =
    -\frac{w}{\eta_y}
    \left(
    \frac23 \mathcal{K} H_d^2+w
    \right),
\end{equation}
where all quantities are evaluated along $x=0$. Introducing
$f(y)=H_d^2(0,y)$,
the equation can be recast in the linear form 
$
    f'(y)+P(y)f(y)+Q(y)=0,
$
with
$
    P(y)=\frac23\frac{w\mathcal K}{\eta_y},
    Q(y)=\frac{w^2}{\eta_y}.
$
The general solution reads
\begin{equation}
    f(y)
    =
    \exp\!\left(
    -\int_{y_0}^{y} P(u)\,\mathrm{d}u
    \right)
    \left[
    f_0
    -
    \int_{y_0}^{y}
    \exp\!\left(
    \int_{y_0}^{s} P(u)\,\mathrm{d}u
    \right)
    Q(s)\,\mathrm{d}s
    \right],
\end{equation}
It can be demonstrated that the constant $f_0$ must vanish, as shown in the next section. We can then rewrite the solution as follows by changing the integral bounds:
\begin{align}
f(y)
&=
-\exp\!\left(
-\int_{y_0}^{y} P(u)\,\mathrm{d}u
\right)
\int_{y_0}^{y}
\exp\!\left(
\int_{y_0}^{s} P(u)\,\mathrm{d}u
\right)
Q(s)\,\mathrm{d}s
\nonumber\\
&=
\int_{y}^{y_0}
\exp\!\left[
-\int_{y_0}^{y}P(u)\,\mathrm{d}u
+
\int_{y_0}^{s}P(u)\,\mathrm{d}u
\right]
Q(s)\,\mathrm{d}s
\nonumber\\
&=
\int_{y}^{y_0}
\exp\!\left[
-\left(
\int_0^yP(u)\,\mathrm{d}u
-
\int_0^{y_0}P(u)\,\mathrm{d}u
\right)
+
\left(
\int_0^sP(u)\,\mathrm{d}u
-
\int_0^{y_0}P(u)\,\mathrm{d}u
\right)
\right]
Q(s)\,\mathrm{d}s
\nonumber\\
&=
\int_{y}^{y_0}
\exp\!\left[
-\int_0^yP(u)\,\mathrm{d}u
+
\int_0^sP(u)\,\mathrm{d}u
\right]
Q(s)\,\mathrm{d}s
\nonumber\\
&=
\exp\!\left(
-\int_0^yP(u)\,\mathrm{d}u
\right)
\int_{y}^{y_0}
\exp\!\left(
\int_0^sP(u)\,\mathrm{d}u
\right)
Q(s)\,\mathrm{d}s .
\end{align}
Substituting the expressions of $f$, $P$, and $Q$, we obtain
\begin{equation}
H_d^2(0,y)
=
\exp\left[
-\frac{2}{3}
\int_0^y
\frac{w\mathcal K}{\eta_y}\,\mathrm{d}u
\right]
\int_y^{y_0}
\exp\left[
\frac{2}{3}
\int_0^s
\frac{w\mathcal K}{\eta_y}\,\mathrm{d}u
\right]
\frac{w^2}{\eta_y}\,\mathrm{d}s .
\end{equation}

We now isolate the logarithmic singular behavior close to the saddle. Along $x=0$, for the saddles considered here,
$
    \eta_y\sim s,
    w\to1, \, \mathcal{K} \sim 0,\,
    \text{as }s\to0.
$
Moreover, the exponential factor inside the integral tends to one. Therefore,
\begin{equation}
\exp\left[
\frac{2}{3}
\int_0^s
\frac{w\mathcal K}{\eta_y}\,\mathrm{d}u
\right]
\frac{w^2}{\eta_y}
\sim
\frac{1}{s},
\qquad
s\to0.
\end{equation}
This $1/s$ term is the only source of the logarithmic divergence. We therefore add and subtract it inside the integral:
\begin{align}
H_d^2(0,y)
=&
\exp\left[
-\frac{2}{3}
\int_0^y
\frac{w\mathcal K}{\eta_y}\,\mathrm{d}u
\right]
\Bigg[
\int_y^{y_0}
\left\{
\exp\left[
\frac{2}{3}
\int_0^s
\frac{w\mathcal K}{\eta_y}\,\mathrm{d}u
\right]
\frac{w^2}{\eta_y}
-
\frac{1}{s}
\right\}
\mathrm{d}s
+
\int_y^{y_0}\frac{1}{s}\,\mathrm{d}s
\Bigg]
\nonumber\\
=&
\exp\left[
-\frac{2}{3}
\int_0^y
\frac{w\mathcal K}{\eta_y}\,\mathrm{d}u
\right]
\Bigg[
\int_y^{y_0}
\left\{
\exp\left[
\frac{2}{3}
\int_0^s
\frac{w\mathcal K}{\eta_y}\,\mathrm{d}u
\right]
\frac{w^2}{\eta_y}
-
\frac{1}{s}
\right\}
\mathrm{d}s
+
\ln\left|\frac{y_0}{y}\right|
\Bigg].
\end{align}
The expression inside braces is now regular at $s=0$, because the singular $1/s$ contribution has been removed. Therefore, in the limit $y\to0$, the lower bound of this regularized integral can be replaced by $0$. Moreover, the exponential prefactor tends to one. Hence,
\begin{equation}
H_d^2(0,y\sim0)
=
\int_0^{y_0}
\left\{
\exp\left[
\frac{2}{3}
\int_0^s
\frac{w\mathcal K}{\eta_y}\,\mathrm{d}u
\right]
\frac{w^2}{\eta_y}
-
\frac{1}{s}
\right\}
\mathrm{d}s
+
\ln|y_0|
-
\ln|y|.
\end{equation}
We identify all finite terms independent of $y$ with the constant $C$, namely
\begin{equation}
C
=
\int_0^{y_0}
\left\{
\exp\left[
\frac{2}{3}
\int_0^s
\frac{w\mathcal K}{\eta_y}\,\mathrm{d}u
\right]
\frac{w^2}{\eta_y}
-
\frac{1}{s}
\right\}
\mathrm{d}s
+
\ln|y_0|.
\end{equation}
Thus, close to the saddle,
$
H_d^2(0,y\sim0)=C-\ln|y|.
$
The expression above can be directly evaluated if $y=y_0$ corresponds to a pinned contact line. If $y_0$ is instead a draining maximum (pole), the integral inside the exponential could be singular as $s\to y_0$ since
$
    \eta_y\sim \eta_{yy}|_{y_0}(s-y_0).
$
We now verify that it is bounded instead. We recall the curvature at the draining maximum,
$
    \mathcal K_p=\mathcal K(0,y_0).
$
Near $s=y_0$, one has
\begin{equation}
    \frac{w^2}{\eta_y}
    \sim
    \frac{1}{\eta_{yy}|_{y_0}(s-y_0)}.
\end{equation}
We verify that this singularity is resolved through the exponential term in the nested integral.
Inside the $u$-integral, the singular behavior as $u\to y_0$ is
\begin{equation}
\frac{w\mathcal K}{\eta_y}
\sim
\frac{\mathcal K_p}{\eta_{yy}|_{y_0}(u-y_0)} .
\end{equation}
We therefore add and subtract this singular part inside this integral:
\begin{align}
\exp\left[
\frac{2}{3}
\int_0^s
\frac{w\mathcal K}{\eta_y}\,\mathrm{d}u
\right]
&=
\exp\left[
\frac{2}{3}
\int_0^s
\left(
\frac{w\mathcal K}{\eta_y}
-
\frac{\mathcal K_p}{\eta_{yy}|_{y_0}(u-y_0)}
+
\frac{\mathcal K_p}{\eta_{yy}|_{y_0}(u-y_0)}
\right)
\mathrm{d}u
\right].
\end{align}
We then separate the two contributions in the exponential:
\begin{equation}
\exp\left[
\frac{2}{3}
\int_0^s
\frac{w\mathcal K}{\eta_y}\,\mathrm{d}u
\right]
=
\exp\left[
\frac{2}{3}
\int_0^s
\left(
\frac{w\mathcal K}{\eta_y}
-
\frac{\mathcal K_p}{\eta_{yy}|_{y_0}(u-y_0)}
\right)
\mathrm{d}u
\right]
\nonumber
\exp\left[
\frac{2}{3}
\int_0^s
\frac{\mathcal K_p}{\eta_{yy}|_{y_0}(u-y_0)}
\,\mathrm{d}u
\right].
\end{equation}
The second exponential can be evaluated explicitly:
\begin{equation}
\exp\left[
\frac{2}{3}
\int_0^s
\frac{\mathcal K_p}{\eta_{yy}|_{y_0}(u-y_0)}
\,\mathrm{d}u
\right]
=
\exp\left[
\frac{2\mathcal K_p}{3\eta_{yy}|_{y_0}}
\int_0^s
\frac{\mathrm{d}u}{u-y_0}
\right]
\nonumber=
\exp\left[
\frac{2\mathcal K_p}{3\eta_{yy}|_{y_0}}
\ln\left(\frac{y_0-s}{y_0}\right)
\right]
\nonumber=
\left(
\frac{y_0-s}{y_0}
\right)^{
2\mathcal K_p/(3\eta_{yy}|_{y_0})
}.
\end{equation}
Therefore,
\begin{equation}
\exp\left[
\frac{2}{3}
\int_0^s
\frac{w\mathcal K}{\eta_y}\,\mathrm{d}u
\right]
=
\exp\left[
\frac{2}{3}
\int_0^s
\left(
\frac{w\mathcal K}{\eta_y}
-
\frac{\mathcal K_p}{\eta_{yy}|_{y_0}(u-y_0)}
\right)
\mathrm{d}u
\right]
\left(
\frac{y_0-s}{y_0}
\right)^{
2\mathcal K_p/(3\eta_{yy}|_{y_0})
}.
\end{equation}
Since both $\mathcal K_p$ and $\eta_{yy}|_{y_0}$ are negative at a draining maximum, the exponent
$
    \frac{2\mathcal K_p}{3\eta_{yy}|_{y_0}}
$
is positive. This factor regularizes the behavior of the integrand in the dummy variable $s$ at $s=y_0$ and allows for explicit numerical integration.
In the draining-pole case, the constant $C$ is thus evaluated using the regularized expression
\begin{equation}
C
=
\int_0^{y_0}
\left\{
\exp\left[
\frac{2}{3}
\int_0^s
\left\{
\frac{w\mathcal K}{\eta_y}
-
\frac{\mathcal K_p}
{\eta_{yy}|_{y_0}(u-y_0)}
\right\}
\mathrm{d}u
\right]
\left(
\frac{y_0-s}{y_0}
\right)^{
2\mathcal K_p/(3\eta_{yy}|_{y_0})
}
\frac{w^2}{\eta_y}
-
\frac{1}{s}
\right\}
\mathrm{d}s
+
\ln|y_0|.
\end{equation}

For the periodic saddle topography from the main text, numerical evaluation of this expression gives
$
    C=1.7949\cdots .
$

\subsection{Constant of integration $f_0$}

We now show that regularity at a draining maximum (pole) imposes the vanishing of the
homogeneous integration constant. Let the draining pole be located at
$x=0$, $y=y_0$, and define
\begin{equation}
    \beta=\frac{2\mathcal K_p}{3\eta_{yy}|_{y_0}} .
\end{equation}
Near the draining pole,
\begin{equation}
    \eta_y\sim \eta_{yy}|_{y_0}(y-y_0),
    \qquad
    \frac{w\mathcal K}{\eta_y}
    \sim
    \frac{\mathcal K_p}{\eta_{yy}|_{y_0}(y-y_0)} .
\end{equation}
The term proportional to the integration constant is
\begin{equation}
    f_{\mathrm h}(y)
    =
    f_0
    \exp\left(
    -\int_{0}^{y}P(u)\,\mathrm{d}u
    \right).
\end{equation}
Near the draining maximum,
\begin{equation}
P(y)\sim \frac{\beta}{y-y_0},
\qquad
\beta=\frac{2\mathcal K_p}{3\eta_{yy}|_{y_0}}>0.
\end{equation}
Hence
\begin{equation}
    f_{\mathrm h}(y)
    \sim
    f_0
    \left(
    \frac{y_0-y}{y_0}
    \right)^{-\beta}.
\end{equation}
This term diverges as $y\to y_0$ unless $f_0=0$. Therefore regularity at the draining maximum imposes $f_0=0$.

\section{Systematic metric expansion close to the saddle - periodic saddle case}

We start from the full model with negligible capillary effects:
\begin{equation}
(1-\mathcal{K}h+\mathcal{G}h^{2})\frac{\partial h}{\partial t}
+\frac{1}{3} \nabla \cdot\left[h^{3} \left({ \boldsymbol{g}}_{t}-  h\left(\mathcal{K} \mathbb{I}+\frac{1}{2} \mathbb{K}\right) \cdot {\boldsymbol{g}}_{t}+  {\mathrm{g}}_{3} \nabla h\right)\right]=0
\end{equation}
and introduce a small parameter $\delta=t^{-1/2} \ll 1$ such that $h = t^{-1/2}\,\mathcal{H}=  \delta \mathcal{H}$:
\begin{equation}
-\delta^3(1-\delta\mathcal{K}\mathcal{H}+\delta^2\mathcal{G}\mathcal{H}^{2})\frac{\mathcal{H}}{2}+\frac{\delta^3}{3} \nabla \cdot\left[\mathcal{H}^{3} \left({ \boldsymbol{g}}_{t}-  \delta \mathcal{H}\left(\mathcal{K} \mathbb{I}+\frac{1}{2} \mathbb{K}\right) \cdot {\boldsymbol{g}}_{t}+  \delta {\mathrm{g}}_{3} \nabla \mathcal{H}\right)\right]=0.
\end{equation}

By simplifying the common term $\delta^3$, at $\mathcal{O}(1)$ one obtains:
\begin{equation}
-\mathcal{H}/2+\frac{1}{3} \nabla \cdot\left[\mathcal{H}^{3} \boldsymbol{g}_{t}\right]=0,
\label{eq:lubr_compl_outer}
\end{equation}
i.e., the drainage problem that describes the solution far from the saddle. As shown in the main text, this equation is singular at the saddle, due to the absence of hydrostatic effects.

The problem close to the considered saddle can be obtained as follows. We develop the metric quantities with a Taylor series around $(x=0,y=0)$ for $\eta(x,y)=1-\cos y -x^2/2$:
\begin{equation}
\begin{gathered}
w = 1+\frac{x^2+y^2}{2}+O(r^4), \qquad
\boldsymbol{g}_{t} = x\boldsymbol{e}_x-y\boldsymbol{e}_y+O(r^3), \qquad
g_3=-1+O(r^2), \qquad \mathcal{G}=-1 +2x^2+\frac52 y^2 +O(r^4) \\
\mathcal{K}=x^2-\frac{3}{2}y^2+O(r^4), \qquad
\partial_x \mathcal{K} = 2x +O(r^3), \qquad
\partial_y \mathcal{K} = -3y +O(r^3), \qquad \partial_x \mathcal G
= 4x+O(r^3),  \qquad
\partial_y \mathcal G
=
5y+O(r^3),\\
\mathbb{G}^{\{i j\}}
=
\mathbb{I}+\begin{pmatrix}
-x^2 & xy\\
xy & -y^2
\end{pmatrix}
+O(r^4), \qquad
\mathbb{K}=
\begin{pmatrix}
-1&0\\
0&1
\end{pmatrix}
+O(r^2),
\end{gathered}
\end{equation}
where $r^2=x^2+y^2$. We introduce $\nabla=\nabla_{xy}:=[\partial_x,\partial_y]$.
Upon substitution, one obtains, up to $\mathcal{O}(r)$ in the divergence:
\begin{equation}
- \frac{ \mathcal{H}}{2}+\nabla_{xy} \cdot\left[\frac{\mathcal{H}^3}{3}\left(x\boldsymbol{e}_x-y\boldsymbol{e}_y+ \delta \mathcal{H}\left(\frac{x}{2}\boldsymbol{e}_x+\frac{y}{2}\boldsymbol{e}_y\right)-\delta {\nabla}_{xy} \mathcal{H}\right)\right]=0.
\label{eq:local_hydr}
\end{equation}
This is a {local Taylor expansion} near the saddle. The term $-\delta\nabla_{xy}\mathcal{H}$ is the classical hydrostatic term. The term $\delta \mathcal{H}(x\mathbf e_x+y\mathbf e_y)/2$ is the leading local curvature-induced hydrostatic correction.}

To obtain the equation, we ``stretch'' along the $y$ direction, consistently with the observed divergence for $y \rightarrow 0$ at $x=0$, i.e. $y  = \delta^{\alpha} \zeta$.
Under the stretching, the transverse drainage term scales as $O(\delta^\alpha)$, the classical hydrostatic term scales as $O(\delta^{1-\alpha})$, and the curvature-induced hydrostatic correction scales as $O(\delta^{1+\alpha})$:
\[
\mathcal{H}^3 y\sim O(\delta^\alpha),\qquad
\delta\,\mathcal{H}^3\partial_y \mathcal{H}\sim O(\delta^{1-\alpha}),\qquad
\delta\,\mathcal{H}^4 y\sim O(\delta^{1+\alpha}).
\]
We obtain
\begin{equation}
-\mathcal{H}/2 +\frac{1}{3} \left[\mathcal{H}^3\left(x + \delta \mathcal{H}\ \frac{x}{2}-\delta {\partial}_{x} \mathcal{H}\right)\right]_x+
\frac{1}{3\delta^\alpha} \left[\mathcal{H}^3\left(-{\delta^\alpha}\zeta + \delta^{1+\alpha} \mathcal{H} \frac{\zeta}{2} - \delta^{1-\alpha}\partial_\zeta\mathcal{H}\right)\right]_\zeta=0.
\label{eq:stretch_hydr}
\end{equation}

The asymptotically relevant dominant balance is obtained by balancing the drainage ${\delta^\alpha}\zeta$ and hydrostatic $\delta^{1-\alpha} \mathcal{H}_\zeta$ terms along the $y$ direction, leading to 
$
\delta^\alpha \sim \delta^{1-\alpha} 
\rightarrow \alpha = 1/2,
$
that gives $\mathcal{O}(1)$ terms after simplification with term $\delta^{-\alpha}$ outside of the divergence. Eventually, we obtain the following equation, at $\mathcal{O}(1)$:
\begin{equation}
    \partial_{\zeta \zeta} \mathcal{H} \mathcal{H}^3 {-}3 x \mathcal{H}^2 \partial_x \mathcal{H} +3 \mathcal{H}^2 \partial_{\zeta} \mathcal{H} \left( \partial_{\zeta} \mathcal{H}+\zeta\right) +\frac{3 \mathcal{H}}{2}=0.
\label{eq:inner_PDE_hydr}
\end{equation}
Although the width is fixed by the local balance, the saddle problem is not fully self-similar because its ansatz satisfies matching with a solution that retains the explicit dependence on the spatially varying logarithmic profile.
We note that the problem after the close-to-the-saddle Taylor expansion and dominant balance coincides with the one obtained by a priori assuming a flat substrate, namely neglecting all metric and curvature effects apart from tangential gravity, as done in the main text. 
Since the equation is of first order along $x$, we restrict our PDE to the line $x=0$, which is also a symmetry plane for $\mathcal{H}$, leading to the final ODE:
\begin{equation}
    \partial_{\zeta \zeta} \mathcal{H} \mathcal{H}^3  +3 \mathcal{H}^2 \partial_{\zeta} \mathcal{H} \left( \partial_{\zeta} \mathcal{H}+\zeta\right) +\frac{3 \mathcal{H}}{2}=0,
\label{eq:inner_ODE_hydr}
\end{equation}
subject to the symmetry condition $\mathcal{H}'(0)=0$ and to the matching with the drainage profile, as detailed in the main.

\subsection{Capillary effects}

For the periodic saddle, i.e., the substrate of elevation 
$
{\eta}({x},{y})=1-\cos ({y}) - {x}^2/2,
$
it is observed that the presence of capillary effects (i.e., a finite Bond number $Bo=\rho g L^2/\gamma$) lets the theoretical profiles deviate from the hydrostatic ridge. To rationalize this, we consider the full model with capillary effects,
\begin{equation}
\begin{aligned}
(1-\mathcal{K}h+\mathcal{G}h^{2})\frac{\partial h}{\partial t}&+\frac{1}{3Bo} \nabla \cdot\left[h^{3} \left( \nabla \tilde{\kappa}-\frac{1}{2} h(2\mathcal{K} \mathbb{I}-\mathbb{K}) \cdot \nabla \mathcal{K}\right)\right] \\
&+\frac{1}{3} \nabla \cdot\left[h^{3} \left({ \boldsymbol{g}}_{t}-  h\left(\mathcal{K} \mathbb{I}+\frac{1}{2} \mathbb{K}\right) \cdot {\boldsymbol{g}}_{t}+  {\mathrm{g}}_{3} \nabla h\right)\right]=0,
\label{eq:lubr_compl_cap}
\end{aligned}
\end{equation}
and expand it close to the saddle, where also
\[
\tilde\kappa=\mathcal K+(\mathcal K^2-2\mathcal G)h+\nabla^2 h
=\left( x^2-\frac32 y^2\right)+(2-4x^2-5y^2)h+\nabla^2_{xy}h+O(r^2),
\]
As before, we assume $h=\delta \mathcal{H} = t^{-1/2} \mathcal{H} +{o}(t^{-1/2})$ and, at $\mathcal{O}(r)$ in the divergence terms, we have (see also next Section):
\begin{multline}
- \frac{ \mathcal{H}}{2}
+\frac{1}{Bo}\nabla_{xy} \cdot\left(\frac{\mathcal{H}^3}{3}\left( 2x\mathbf{e}_x-3y\mathbf{e}_y +\nabla_{xy} \left( (2-4x^2-5y^2)\delta \mathcal{H} +\delta \nabla_{xy}^2 \mathcal{H} \right) {+} \frac{\delta}{2}\mathcal{H}\left(-\mathcal{K}_x \mathbf{e}_x +\mathcal{K}_y \mathbf{e}_y \right)\right)\right) \\
+\nabla_{xy} \cdot\left(\frac{\mathcal{H}^3}{3}\left( x\boldsymbol{e}_x-y\boldsymbol{e}_y +\delta \mathcal{H}\left(\frac{x}{2}\boldsymbol{e}_x+\frac{y}{2}\boldsymbol{e}_y\right)-\delta {\nabla}_{xy} \mathcal{H}\right)\right)=0.
\label{eq:4}
\end{multline}

\begin{figure}[t]
    \centering
\includegraphics[width=0.5\linewidth]{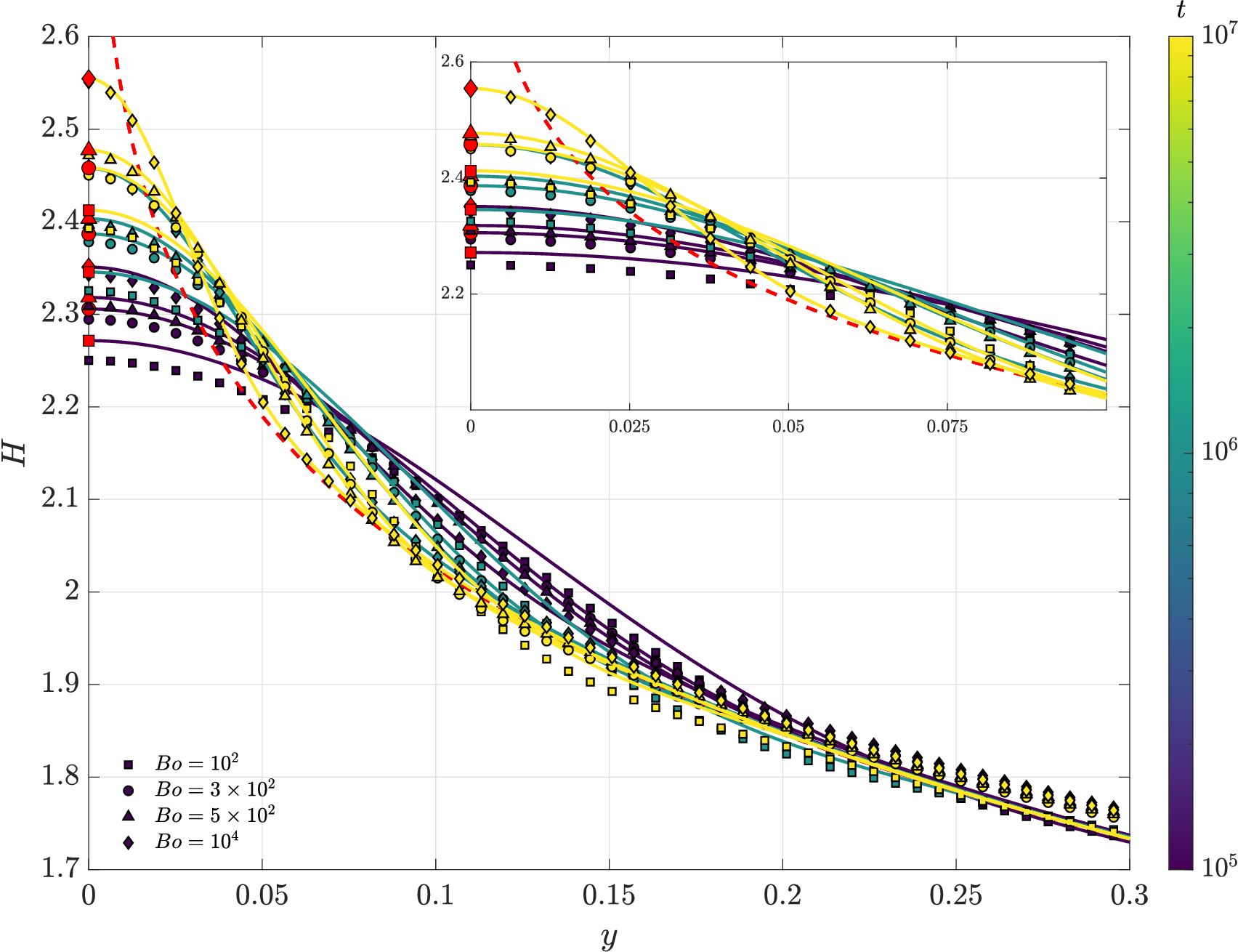}
    \caption{Numerical profiles of $\mathcal{H}=t^{1/2}h$, with initial condition $h_0=0.1$, obtained from the full model close to the saddle for varying Bond number (symbols), increasing time (colormap) and comparison with the theoretical result for varying Bond (lines). The zoom in the inset helps distinguish the curves and the agreement across diverse times and Bond numbers. As the Bond number increases, the ridge maximum height increases. }
    \label{fig:bond}
\end{figure}

The balance for the stretched variable $\zeta=y/\delta^\alpha$ now depends on the order of magnitude of the Bond number. If $Bo \gg t^{1/2}=\delta^{-1}$, the term disappears from the previous equation, giving the previously derived pure hydrostatic problem.
Capillary effects become more important as time increases since the saddle layer narrows, increasing the order of magnitude of the high derivative terms related to the film curvature. For large-enough time, capillary and hydrostatic term are of the same order of magnitude.

As discussed in the main, $Bo = \mathcal{O}( t^{1/2})=\mathcal{O}(\delta^{-1}) := Bo^*t^{1/2}$. Therefore, the Bond number now takes into account the hydrostatic  thinning with a dynamic length scale $\hat \ell_b/L \sim t^{-1/4}$, i.e.  $Bo = (\hat \ell_b^2/\ell_c^2) t^{1/2}$, and $Bo^*=\hat \ell_b^2/\ell_c^2$ is the Bond number referring to this dynamic length scale. 
Under the combined scaling
\[
y=\delta^{1/2}\zeta,\qquad Bo^{-1}=\frac{\delta}{Bo^*},
\]
the different capillary contributions in Eq.~\eqref{eq:4} scale, after the divergence, as
\[
\frac{1}{Bo}\nabla\cdot\!\left(\mathcal{H}^3\nabla_{xy}\mathcal K\right)=O(\delta),\qquad
\frac{1}{Bo}\nabla\cdot\!\left(\mathcal{H}^3\,2\delta\nabla_{xy}\mathcal{H}\right)=O(\delta),
\]
\[
\frac{1}{Bo}\nabla\cdot\!\left(\mathcal{H}^3\,\delta\nabla_{xy}\nabla_{xy}^2\mathcal{H}\right)=O(1),
\qquad
\frac{1}{Bo}\nabla\cdot\!\left[\delta \mathcal{H}^4(2\mathcal K I-\mathbb K)\cdot\nabla_{xy}\mathcal K\right]=O(\delta^2).
\]

As a consequence, the scaling is the same as before, i.e. $\zeta=yt^{1/4}$ and, at $\mathcal{O}(1)$, only the highest derivatives of capillarity balance with hydrostatics and drainage:
\begin{equation}
- \frac{ \mathcal{H}}{2}+\frac{1}{Bo^*}\partial_\zeta \left(\frac{\mathcal{H}^3}{3}\left(\partial_\zeta^3 \mathcal{H}\right)\right) +\partial_x \left(\frac{\mathcal{H}^3}{3} x \right) +\partial_\zeta \left(\frac{\mathcal{H}^3}{3} \left(-\zeta -\partial_\zeta \mathcal{H} \right) \right) =0,
\label{eq:4_innercap}
\end{equation}
i.e., along $x=0$,
\begin{equation}
  \mathcal{H}'' \mathcal{H}^2  -\frac{1}{Bo^*}\left(3 \mathcal{H} \mathcal{H}' \mathcal{H}''' + \mathcal{H}^2 \mathcal{H}^{\mathrm{IV}}\right) +3 \mathcal{H} \mathcal{H}' \left( \mathcal{H}'+\zeta\right)+\frac{3}{2} =0,
\label{eq:innercap_ODE}
\end{equation}
subject to the symmetry conditions $\mathcal{H}'(0)=\mathcal{H}'''(0)=0$ and to the same matching as the pure hydrostatic problem, for the value and first derivative of the logarithmic drainage profile.
Consequently, 
the leading problem 
is again equivalent to the equation obtained by assuming a flat substrate from the beginning, and hence is generalizable to any substrate presenting a saddle of hyperbolic form, i.e. $\eta \sim y^2-x^2$.
A comparison between the solution of problem Eq.~\eqref{eq:innercap_ODE} with numerical simulation profiles is reported in Fig.~\ref{fig:bond}, with a good agreement.

\subsection{Local expansion of the Laplace--Beltrami operator at the saddle}

We here clarify how the Laplacian term does not produce any additional term more than the Cartesian form, before the dominant balance.
The Laplace--Beltrami operator acting on a scalar field $f(x,y)$ reads
\[
\nabla^2 f=\frac{1}{w}\partial_i\!\left(w\,\mathbb G^{\{ij\}}\partial_j f\right),
\]
We now substitute the Taylor expansion into the Laplace--Beltrami operator:
\[
\nabla^2 f
=
\frac{1}{w}
\left[
\partial_x\!\left(
\left(1-\frac{x^2}{2}+\frac{y^2}{2}\right)\partial_x f +xy\,\partial_y f
\right)
+
\partial_y\!\left(
xy\,\partial_x f+\left(1+\frac{x^2}{2}-\frac{y^2}{2}\right)\partial_y f
\right)
\right]
+O(r^2),
\]

Expanding the derivatives yields
\[
\begin{aligned}
\nabla^2 f
&=
\frac{1}{w}\Bigg[
\left(1-\frac{x^2}{2}+\frac{y^2}{2}\right)\partial_{xx} f
-x\partial_x f+y\partial_y f
+2xy\,\partial_{xy} f
+x\partial_x f
+\left(1+\frac{x^2}{2}-\frac{y^2}{2}\right)\partial_{yy} f
-y\partial_{y} f
\Bigg]
+O(r^2).
\end{aligned}
\]
Since the linear first-derivative terms cancel exactly, multiplication by $w^{-1}=1+O(r^2)$ does not generate linear terms and we can neglect all other quadratic terms. Therefore, the Laplace-Beltrami operator coincides with the Cartesian Laplacian up to linear order near the saddle. In conclusion, no terms of order $O(x)$ or $O(y)$ stem from metric, in the highest-derivative capillary contribution.

\section{The hyperbolic paraboloid problem }

Characteristic lines are such that $x y =\text{const.}:= A$. The drainage problem thus becomes
\begin{equation}
\frac{\mathrm{d} H_d^2}{\mathrm{d} y}=-\dfrac{2}{3}\dfrac{A^2-y^4}{y(A^2+y^2+y^4)}H_d^2-\dfrac{A^2+y^2+y^4}{y^3}\label{eq:ode-hyper-para}
\end{equation}

\begin{figure}[t]
    \begin{subfigure}[t]{0.22\textwidth}
    \caption{\raggedright}
    \includegraphics[width=\linewidth]{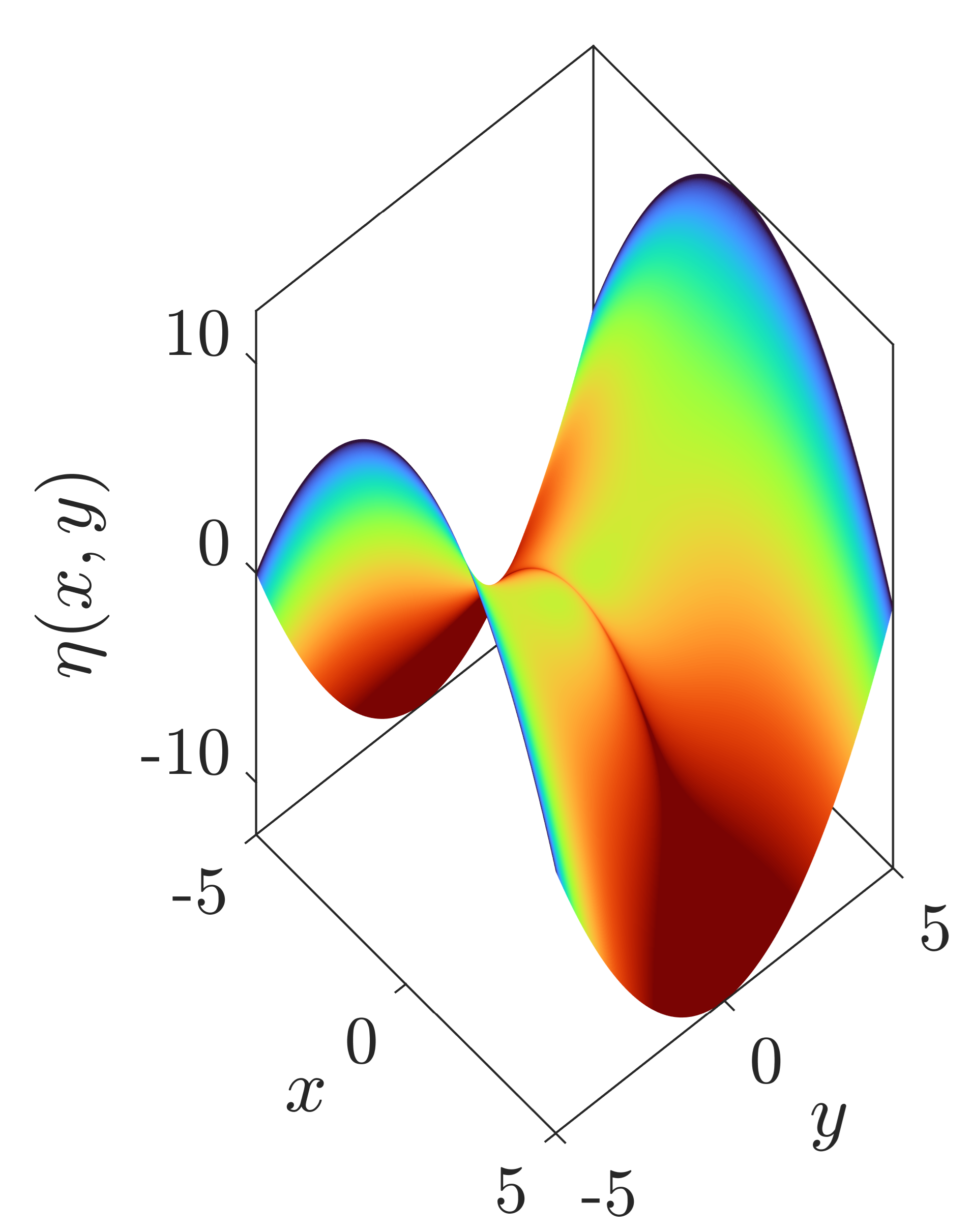}
\end{subfigure}
\hfill
\begin{subfigure}[t]{0.4\textwidth}
\caption{\raggedright}    \includegraphics[width=\linewidth]{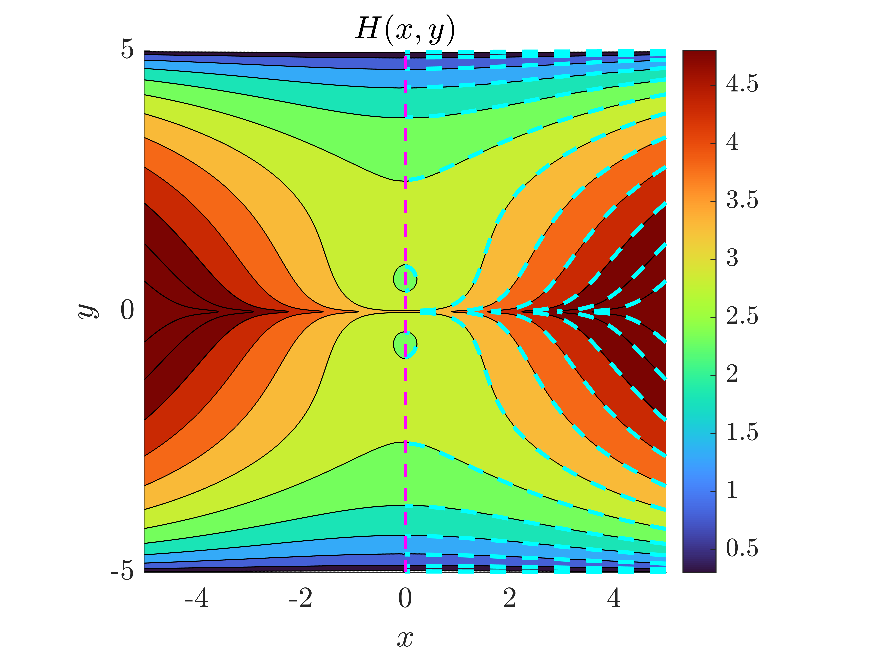}
    
\end{subfigure}
\hfill
\begin{subfigure}[t]{0.36\textwidth}
    \caption{\raggedright}
    \includegraphics[width=\linewidth]{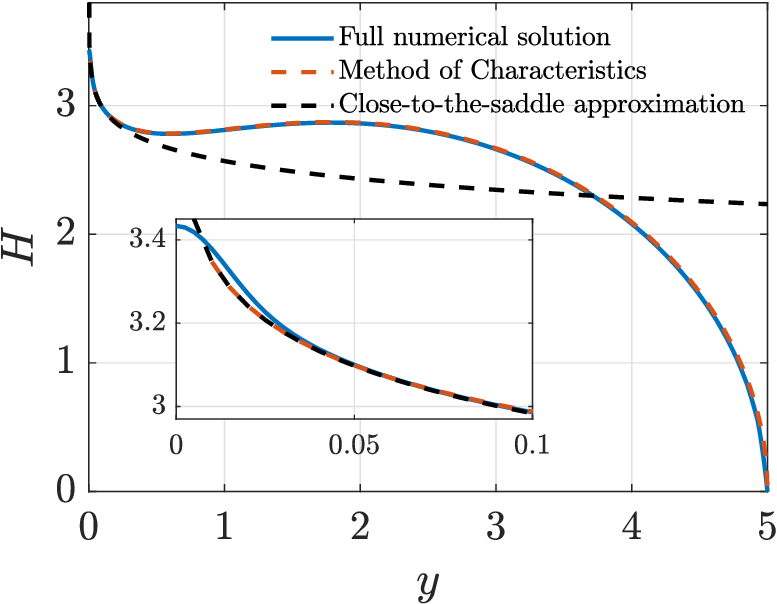}
    
\end{subfigure}
\hfill
    \caption{Hyperbolic paraboloid problem with a pinned contact line at $y=5$. (a,b) Numerical results ($Bo \rightarrow \infty$) for $h_0:=h(x,y,0)=0.1$ and $t=10^6$: (a) Colormaps of $H(x,y)$ on a surface plot of the substrate height. (b) Colormaps and isocontours (black) of $H(x,y)$. Cyan dashed lines: same isocontours for the drainage solution, reported only for $x>0$ because of symmetry with respect to $x=0$ (magenta dashed line). (c) Thickness $H$ as a function of $y$, for $x=0$ for $t=10^{8}$. The orange dashed line is the $\delta=0$ solution and the black dash-dotted line its asymptotic behavior close to the saddle. In the inset, a zoom close to the saddle. }
    \label{fig:xysaddle}
\end{figure}

In our case, we impose $H_d(x_0,y_0)=0$, with $y_0=5$ and $x_0=A/y_0$ with varying $A$, delimiting the wetted area. 
This problem is solved in MATLAB via the routine \textit{ode45}, with a relative and absolute tolerance of $10^{-7}$ and $10^{-10}$, respectively.
Eq.~\eqref{eq:ode-hyper-para} can be explicitly integrated for $x=0$, which implies $A=0$,
\begin{equation}
H_d^2(x=0,y)=(1+y^2)^{1/3}\int_y^{y_0}\dfrac{(1+y'^2)^{2/3}}{y'}\mathrm{d}y',
\label{eq:ode-hyper-para-converg-sol}
\end{equation}
which leads to $H_d(x=0,y\rightarrow 0)=\sqrt{C-\ln(y)}$ with \[C=\ln|y_0|+\int_0^{y_0}\dfrac{(1+y^2)^{2/3}-1}{y}\mathrm{d}y=6.5984\cdots \text{ for } y_0=5,\]
We compare this drainage solution with the full model with $Bo\rightarrow \infty$, which is obtained by changing the substrate expression in our formulation and by imposing a zero thickness boundary condition (pinned contact line) at $y=5$, with a sponge method to relax the thickness to zero beyond $x=6$, while keeping tolerances and mesh the same as the periodic saddle case.
As shown in Fig.~\ref{fig:xysaddle}, the same ridge along $y=0$ as in the periodic saddle case is observed, with the full numerical solution that agrees well with the drainage solution. The same structure is observed for this substrate, confirming that the local saddle topology drives the ridge formation rather than the whole substrate topology.

\bibliography{biblio}